\documentclass[iop,revtex4]{emulateapj}

\begin{document}
 
\newcommand{\kms}{km s$^{-1}\;$}
\newcommand{\msun}{M_{\odot}}
\newcommand{\rsun}{R_{\odot}}
\newcommand{\teff}{T_{\rm eff}}
\newcommand{\logg}{\log g}
\newcommand{\kep}{{\it Kepler}~}
\makeatletter
\newcommand{\Rmnum}[1]{\expandafter\@slowromancap\romannumeral #1@}
\newcommand{\rmnum}[1]{\romannumeral #1}
 
\title{The Age and Distance of the {\it Kepler} Open Cluster NGC 6811 from an
  Eclipsing Binary, Turnoff Star Pulsation, and Giant Asteroseismology
\footnote{Based on observations made with with the Hobby-Eberly
    Telescope, which is a joint project of the University of Texas at
    Austin, the Pennsylvania State University, Stanford University,
    Ludwig-Maximilians-Universit\"{a}t M\"{u}nchen, and
    Georg-August-Universit\"{a}t G\"{o}ttingen; and the Nordic Optical
    Telescope, operated by the Nordic Optical Telescope Scientific
    Association at the Observatorio del Roque de los Muchachos, La
    Palma, Spain, of the Instituto de Astrofisica de Canarias}}

\author{Eric L. Sandquist\altaffilmark{1}; J. Jessen-Hansen\altaffilmark{2};
  Matthew D. Shetrone\altaffilmark{3};
  Karsten Brogaard\altaffilmark{2,4}; 
 S$\o$ren Meibom\altaffilmark{5}; Marika Leitner\altaffilmark{6}; Dennis Stello\altaffilmark{7}; Hans Bruntt\altaffilmark{2}; 
Victoria Antoci\altaffilmark{4}; Jerome A. Orosz\altaffilmark{1}; 
Frank Grundahl\altaffilmark{4}; S$\o$ren
  Frandsen\altaffilmark{4}}
\altaffiltext{1}{San Diego State University, Department of Astronomy,
  San Diego, CA, 92182; {\tt esandquist@mail.sdsu.edu,
    jorosz@mail.sdsu.edu}} \altaffiltext{2}{Stellar Astrophysics
  Centre, Department of Physics and Astronomy, Aarhus University, Ny
  Munkegade 120, DK - 8000 Aarhus C, Denmark; {\tt
    jjh@phys.au.dk, kfb@phys.au.dk, bruntt@gmail.com,
    antoci@phys.au.dk, fgj@phys.au.dk, srf@phys.au.dk}}
\altaffiltext{3}{University of Texas, McDonald Observatory, HC75 Box
  1337-L Fort Davis, TX, 79734; {\tt shetrone@astro.as.utexas.edu}}
\altaffiltext{4}{Department of Physics \& Astronomy, University of
  Victoria, P.O. Box 3055, Victoria, BC V8W 3P6, Canada}
\altaffiltext{5}{Harvard-Smithsonian Center for Astrophysics,
  Cambridge, MA 02138; {\tt smeibom@cfa.harvard.edu}}
\altaffiltext{6}{Humboldt State University, Department of Physics \&
  Astronomy, 1 Harpst St., Arcata, CA, 95521; {\tt
    rika.six@gmail.com}} \altaffiltext{7}{Sydney Institute for
  Astronomy (SIfA), School of Physics, University of Sydney, NSW,
  2006, Australia; {\tt stello@physics.usyd.edu.au}}

\begin{abstract}
We present the analysis of an eccentric, partially eclipsing
long-period ($P=19.23$ d) binary system KIC 9777062 that contains
main sequence stars near the turnoff of the intermediate age open cluster NGC
6811. The primary is a metal-lined Am star with a
possible convective blueshift to its radial velocities, and one star (probably
the secondary) is likely to be a $\gamma$ Dor pulsator.
The component masses are $1.603\pm0.006$(stat.)$\pm0.016$(sys.) and
$1.419\pm0.003\pm0.008 \msun$, and the radii are $1.744\pm0.004\pm0.002$ and
$1.544\pm0.002\pm0.002 \rsun$. The isochrone ages of the stars
are mildly inconsistent:
the age from the mass-radius combination for the primary
($1.05\pm0.05\pm0.09$ Gyr, where the last quote was systematic
uncertainty from models and metallicity) is smaller than that from the
secondary ($1.21\pm0.05\pm0.15$ Gyr) and is consistent with the
inference from the color-magnitude diagram ($1.00\pm0.05$
Gyr).

We have improved the measurements of the asteroseismic parameters
$\Delta \nu$ and $\nu_{\rm max}$ for helium-burning stars in the
cluster. The masses of the stars appear to be larger (or alternately,
the radii appear to be smaller) than predicted from isochrones using
the ages derived from the eclipsing stars.

The majority of stars near the cluster turnoff are
pulsating stars: we identify a sample of 28 $\delta$ Sct, 15 $\gamma$
Dor, and 5 hybrid types.  We used the period-luminosity relation for
high-amplitude $\delta$ Sct stars to fit the ensemble of the strongest
frequencies for the cluster members, finding $(m-M)_V =
10.37\pm0.03$. This is larger than most previous determinations,
but smaller than values derived from the eclipsing binary
($10.47\pm0.05$).
\end{abstract}


\section{Introduction}

With the completion of a long stare at its original field, a host of
new astrophysical information is available on the star clusters
observed during the original \kep mission. Variable stars of many
different types were discovered and studied thanks to the
high-precision photometry and the nearly total coverage of the cycles
of even long-period variables. One of the challenges in the aftermath
of the \kep mission continues to be the synthesis of information
derived from large samples of stars.

Our focus in this paper is the sample of stars from one of the open
star clusters in the original \kep field.  The brightest stars of the
open cluster NGC 6811 comprise a rich set of objects with
astrophysical information that has not yet been fully exploited. On
the cool side, there are a group of helium-burning red clump and
asymptotic giant stars with solar-like oscillations previously
detected with the \kep spacecraft
\citep{stelloa,stello,hekker,corsaro}. On the hot side, the bright end
of the main sequence fits neatly within the instability strip, and
$\delta$ Scuti (Sct) and $\gamma$ Doradus (Dor) pulsating stars have
been previously identified \citep{vanc,luo,deboss,uytter}. This part
of the color-magnitude diagram is known to contain chemically peculiar
Am/Fm and Ap stars, although none has previously been identified in
this cluster. With the rich set of information available for these
kinds of stars, astrophysical information from one type of star will
undoubtedly illuminate the others.

The main, but not the only, focus of the present paper is the
brightest known eclipsing binary in the cluster.  The eclipsing binary
KIC 9777062 ($\alpha_{2000} = 19^{h}37^{m}50\fs58$,
$\delta_{2000}=+46\degr35\arcmin23\farcs0$; also star 195 in
\citealt{sanders6811}, and WEBDA 484) under consideration here was
first identified in the \kep Eclipsing Binary Catalog \citep{kebs1}.
\kep observed this system almost continuously during the main mission,
thanks to a special effort by the science team to study clusters (the
\kep Cluster Study; \citealt{meibom11}).  Long-period eclipsing binary
stars are treasures for stellar astrophysicists because they offer the
chance to examine the characteristics of (effectively) isolated stars
with a high degree of precision and accuracy. Mass and radius in
particular can be simultaneously measured to precisions of better than
1\% in many cases \citep{andersen,torres}. When one or both of the
stars in the binary has evolved significantly off of the zero-age main
sequence and approaches the end of core-hydrogen burning, the radius
becomes an age-sensitive quantity. And if the binary is a member of a
star cluster, this gives us a means of tightly constraining the age of
the cluster and testing stellar evolution theory quite strictly
\citep[e.g.][]{brogaard2,brogaard,meibom}.

Variability study of NGC 6811 began with \citet{vanc} and continued
with \citet{rh} and \citet{luo}. Most variable star detections have
been pulsating stars (like $\delta$ Sct) near the turnoff of the
cluster, although two possibly eclipsing variables were discovered by
\citeauthor{vanc}.  \citet{balona} examined variability for three
early A-type stars in NGC 6811, including one that appeared to be a
binary star. 

As part of this paper, we have examined the photometry and membership
of nearly all bright potential members of the cluster. This step is
particularly important for a relatively sparse cluster like NGC 6811
because of the small number of stars in the more advanced stages of
evolution. But even so, high quality photometry for the
color-magnitude diagram (CMD) can nonetheless provide strong
constraints on the age of the cluster and important physical effects
influencing stars near the cluster turnoff.  In \S \ref{obs}, we
describe the observational material we have gathered on the binary KIC
9777062 and other stars near the turnoff of NGC 6811.  In \S
\ref{analy}, we present the analysis of the eclipsing binary KIC
9777062. In the discussion in \S \ref{disc}, we describe the various
observational constraints on the characteristics of NGC 6811,
including the cluster's distance and age.

\subsection{Cluster Chemical Composition and Reddening}\label{chemred}

The chemical composition and reddening values for the cluster are
important for the interpretation of many aspects of the work described
here, so we will first summarize previous results from the literature
and present one new determination of the reddening.

Until very recently, metallicity determinations have been indirect and
not from spectroscopy. \citet{mz13} obtained high-resolution spectra
of 5 cluster giants, and the average metallicity derived from these
stars were [Fe/H]$=0.00\pm0.02$ and $+0.01\pm0.04$ using two different
abundance analysis algorithms. It is important to note that i) neither
of these measurements employed asteroseismic $\logg$ measurements,
and the average deviation of the measured $\logg$ from the
asteroseismic values is about 0.5 dex for the second algorithm, and
ii) the second of these measurements was differential with respect to
solar abundances, whereas the remaining measurements described below
are not.
\citet{mz} used medium-resolution ($R = 25000$) spectroscopy along
with asteroseismic $\logg$ values for the observed giants to find an
average of [Fe/H]$=+0.04\pm0.01$, where the quoted uncertainty does
not include sources of systematic errors. (Three of the five giant
stars they used overlapped with the 2013 sample.) Six cluster giants
also have infrared spectroscopic abundances from the APOGEE survey
\citep[Data Release 12;][]{apogee}, which uses synthetic spectra (employing an
\citealt{asp05} solar abundance mix) as comparisons.  For these stars,
the average metal content is [M/H]$=+0.05\pm0.02$, and iron abundance
is [Fe/H]$=+0.02\pm0.02$.  \citet{occam} derived a value of
[M/H]$=-0.02\pm0.04$ from APOGEE pipeline analysis of 2 members. Even
with all of the caveats, the different spectroscopic measurements
agree surprisingly well.

For our ultimate comparisons with models, however, we need to consider
effects on the total metal mass fraction $Z$ more carefully. The
APOGEE project produces abundance measurements of the CNO elements
that contribute about half of the metal content. For the six measured
giant stars, the average abundances were [C/H]$=-0.12\pm0.02$,
[N/H]$=+0.26\pm0.03$, and [O/H]$=+0.03\pm0.02$. \citet{mz} found solar
abundance ratios (with the exception of an overabundance of Ba) for 31
elements in 5 stars. Although the individual element differences with
the Sun are potentially interesting, the CNO elements taken together
imply only a small modification of the $Z$ relative to the Sun.  We
will return to this discussion in Section \ref{disc}.

Early studies of the cluster reddening used $E(B-V)=0.16$, but more
recent ones have shown evidence for a lower value. The galactic
reddening map of \citet{sfred} indicates that the integrated reddening
along lines of sight through NGC 6811 should be $E(B-V)$ between 0.129
and 0.135, and this should be an upper limit for the cluster. The
cluster reddening has been re-evaluated recently, although all of
these newer studies effectively compare photometric data to isochrones
and are therefore subject to a host of uncertainties, both
  theoretical (differences in the physics in the stellar models, and
  color-temperature transformations) and observational (such as the
  weighting of the contributions from the photometric sample). Three
studies \citep{glush,janes,yontan} used $UBVRI$ photometry to derive
reddenings. \citet{glush} found $E(B-V)=0.12\pm0.02$ from fits of an
older generation of isochrones to their $UBVRI$ photometry in the CMD
and two-color diagrams.
\citet{janes} derived $E(B-V)=0.074\pm0.024$ from a Bayesian
comparison of their $UBVRI$ photometry with Yale-Yonsei and Padova
isochrones in various CMDs.  \citet{yontan} derive
$E(B-V)=0.046\pm0.012$ from a Bayesian comparison of their $UBVRI$
photometry with Padova isochrones in CMDs and two-color diagrams.
\citet{pena} determined $E(B-V)=0.14$ from Str\"{o}mgren photometry,
but their cluster member CMD has a large amount of scatter and they
derive an unrealistically small age (about 190 Myr), so we do not
believe this value is reliable. \citet{mz} quote $E(B-V)=0.05\pm0.02$
based on a comparison with PARSEC isochrones using spectroscopic
$T_{\rm eff}$ and asteroseismic information.

To get an independent measure of the interstellar reddening $E(B-V)$
we exploited the calibration of the interstellar Na D line strength to
$E(B-V)$ by \citet{munari}. We measured the equivalent width $W$ of
the interstellar Na lines from the spectra of five cluster member
giants obtained by \citet{mz}, subtracting in each case the
contribution from the overlapping stellar Na lines by employing a
template spectrum from \citet{coelho} with stellar parameters close to
those measured by \citet{mz}. To be exact, the parameters were $\log
g=3.0$, [Fe/H]$=0.0$, and $\teff=5000$ K. We made use of both the D1
and D2 lines, and the wavelength ranges over which the equivalent
width was measured were $5888.5-5890.8$ \AA\ and $5894.5-5896.9$ \AA,
respectively. We used a factor of 1.1 to estimate $W$(D1) from $W$(D2)
(see \citealt{munari}), and found agreement with the direct estimate
of $W$(D1) for most stars at a level below 0.01 and never worse than
0.02. Averaging the results from the two lines and then from the 10
spectra (two of each star) yielded $W(\mbox{D1})=0.192\pm0.027$. Using
the calibration in table 2 of \citeauthor{munari} this resulted in
$E(B-V)=0.068\pm0.009$ where the uncertainty is the standard deviation
of the mean. However, to account also for the uncertainty in the
calibration we adopt $E(B-V)=0.07\pm0.02$ as our reddening
estimate. We note that \citeauthor{munari} state a much larger
uncertainty on the calibration, but we argue as in \citet{brogaard}
that the scatter around their calibration is very low at $E(B-V)$
values below 0.2 (see lower panel of their figure 4). As additional
evidence, this method was used by \citet{brogaard} to determine the
reddening of the open cluster NGC 6791, producing a value in good
agreement with other estimates. We will use $E(B-V) = 0.07\pm0.02$ as
our primary choice when needed in the analysis below.

\section{Observational Material and Data Reduction}\label{obs}

\subsection{\kep Photometry}\label{kphot}

Long cadence (30 min exposure time) data were recorded for KIC 9777062
during the entire 17 quarter duration of the main \kep mission. In
addition, short cadence (1 min exposure time) observations were taken
during 6 different month-long periods within quarters 4, 10, and 14.

For our analysis we made use of light curves derived from the
Pre-search Data Conditioning (PDC) pipeline \citep{stumpe,smith} from
Data Release 21, but we verified that these were good representations
of the eclipses by comparing to our own reduction of the raw pixel
data. The PDC pipeline cotrends light curves against those of other
stars on the same CCD to identify and remove systematic trends
\citep{kinemuchi}, and also applies corrections to the flux
measurements to account for contamination by other nearby stars.
Although KIC 9777062 is in the field of NGC 6811, the cluster is a
fairly sparse one and KIC 9777062 is well outside the densest parts of
the cluster. As a result, contaminating light from other stars is a
relatively minor consideration: the contamination index (ratio of
contaminating flux to total flux in the photometric aperture) from the
\kep Data Search at the Mikulski Archive for Space Telescopes
\citep[MAST;][]{brown11} ranged from 0.007 to 0.014.

\begin{figure*}
\includegraphics[scale=0.4]{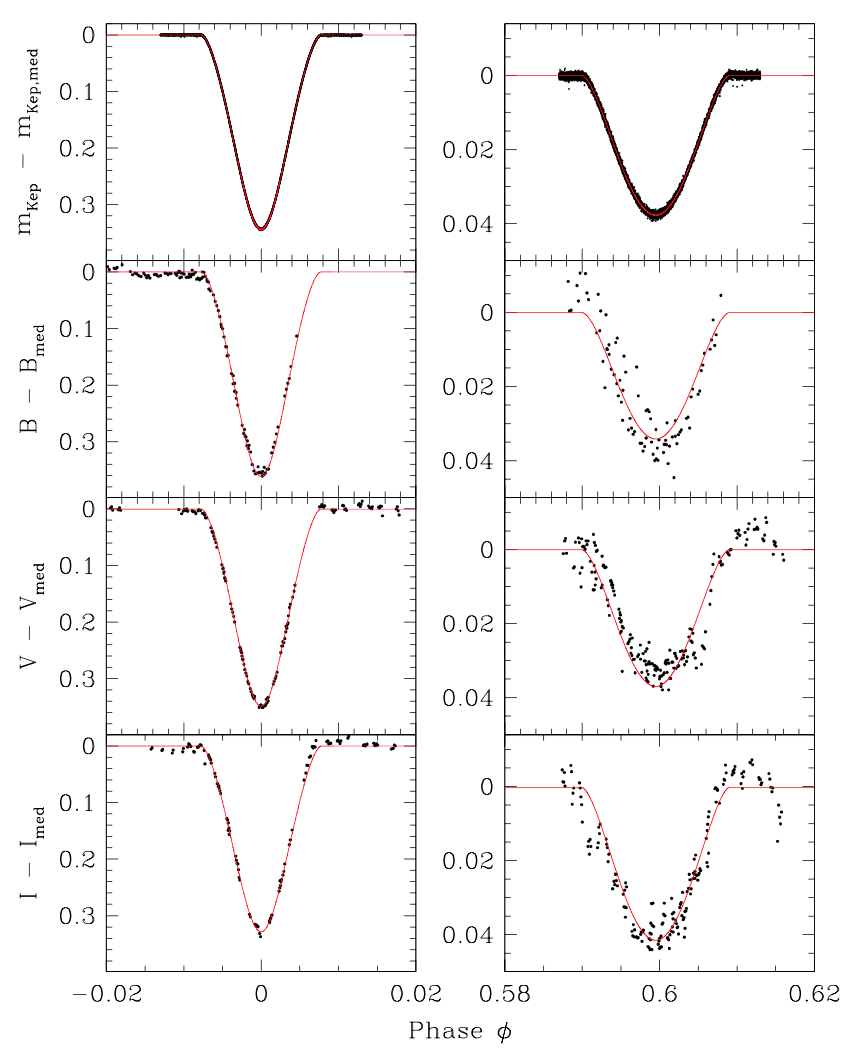}
\caption{{\it Top row:} \kep photometry of the primary and secondary eclipse
  of KIC 9777062. {\it Bottom rows:} Ground-based $BVI_C$ photometry of the
  primary and secondary eclipses. \label{phot}}
\end{figure*}

After some trials, we settled on the use of short cadence data in our
analysis, largely for the greater time resolution. Five of
the six months of short cadence data were also taken during the same
\kep observing season, so that the majority of the data came from
similar parts of the same CCD module, reducing the importance of
instrumental corrections. The corrected \kep eclipse light curves are
shown in Fig. \ref{phot}, with the median out-of-eclipse magnitude
($m_{Kep,med}$) subtracted. The primary and secondary eclipses have
different duration and the phase difference between the eclipses is
about 0.6, both of which are indicators of significant
eccentricity in the orbits.

\subsection{Ground-based Photometry}\label{gbphot}

In order to obtain color information for the binary, we obtained eclipse
photometry in $BVI_c$ using the Mount Laguna Observatory (MLO) 1m
telescope. The observational details are given in Table
\ref{tableobs}. Typical exposure times were 300s in $B$, 180 s in $V$ and 120
s in $I_c$.  The images were processed using standard IRAF\footnote{IRAF is
  distributed by the National Optical Astronomy Observatory, which is operated
  by the Association of Universities for Research in Astronomy, Inc., under
  cooperative agreement with the National Science Foundation.} routines to
apply overscan corrections from each image, to subtract a master bias frame,
and to divide a master flat field frame.
A nonstandard aspect of the image processing involved the correction for a
nonlinearity in the CCD response that was present as a result of improperly set
readout transistor voltages used between December 2008 and
November 2013. The following correction (D. Leonard, private communication)
$ADU_{cor} = ADU \cdot \left[1.01353 - 0.115576 \cdot (ADU / 32767)
  + 0.0296378 \cdot (ADU / 32767)^2\right]$
was applied using the IRAF task {\it irlincor} to pixel counts on flat field
and object images (after overscan and bias subtraction) taken during that time
frame. The nonlinearity had small but noticeable effects on photometric
scatter and trends within nights of data.

\begin{deluxetable}{lccc}
\tablecolumns{4}
\tablewidth{0pc}
\tablecaption{Photometric Observations at Mount Laguna Observatory 1-meter}
\tablehead{
\colhead{UT Date} & \colhead{Filters} & \colhead{mJD\tablenotemark{a} Start} &
\colhead{$N$}}
\startdata
2012 Jun. 2 & $VI_c$ & 6080.71580 & 68,46 \\
2012 Jul. 3 & $VI_c$ & 6111.65676 & 77,75 \\
2012 Jul. 22 & $VI_c$ & 6130.66256 & 80,80 \\
2012 Jul. 30 & $VI_c$ & 6138.66063 & 42,49 \\
2012 Aug. 26 & $BVI_c$ & 6165.72618 & 5,5,5; calibration \\
2013 Jun. 14 & $BI$ & 6457.68656 & 52,28 \\
2013 Jul. 30 & $BI$ & 6503.65531 & 67,26 \\
2013 Aug. 9 & $BVI_c$ & & calibration \\
2014 May 26 & $BV$ & 6803.73965 & 40,21 \\
2014 Jun. 22 & $BV$ & 6830.68880 & 46,27\\%
2014 Jul. 30 & $BV$ & 6868.65135 & 47,48 \\
2014 Aug. 11 & $BV$ & 6880.65445 & 36,40
\enddata
\tablenotetext{a}{mJD = BJD - 2450000}
\label{tableobs}
\end{deluxetable}

The secondary eclipses of the binary are $0.03-0.04$ mag deep, and so
relatively small data reduction issues can have large effects on the
fidelity of the light curve. Our brightness measurements were derived
from aperture photometry using DAOPHOT \citep{daophot}, although we
took several additional steps to increase the precision of the
results. We conducted a curve-of-growth analysis of 12 apertures
photometered per star using DAOGROW \citep{daogrow} in order to
correct all measured stars to a uniform large aperture.  We then
attempted to unify the data for each filter to a consistent zeropoint
by using ensemble photometry \citep{s1082,honey}. This essentially
uses all measured stars on the frame to determine magnitude offsets
resulting from differences in exposure time, airmass, atmospheric
transparency, and the like. Our implementation iteratively fits for
position-dependent corrections that result from variations from
point-spread function across the frame.  These steps each brought
noticeable reductions in the amount of scatter in the light
curves. However, we still found that the shape of the ground-based
light curves of the secondary eclipse did not match what was expected
from the \kep observations. After some investigation, we found that
features in the light curve were correlated with those for the
brightest star near KIC 9777062 in our images. (The star was about 70
pix or $28\arcsec$ distant on the sky, so there was no significant
overlap of the point-spread function when the seeing was generally 4-8
pix FWHM.)  We found that after subtracting the light curve of this
star the out-of-eclipse light levels and the shapes of the primary and
secondary eclipse light curves were much more consistent from night to
night. When more distant stars on the images were tested, we found a
much poorer degree of correlation. We conclude that short length-scale
variations were not being corrected for (and could not be corrected
for, due to lack of star sampling on the image) with whole-image
zeropoint corrections or image-scale position-dependent
corrections. The variations were of around 0.01 mag size, but could
significantly affect the secondary eclipse light curves. We could not
identify any features on our flat field images that could explain the
light curve variations, and they occurred whether dome flats, twilight
flats, or hybrid flats (combining the smoothed large-scale variations
from the twilight flats and small-scale variations from dome flats)
were used.  The ground-based eclipse observations are shown in the
bottom rows of Fig. \ref{phot}.

In addition to the eclipse light curve observations, we took $BVI_C$
images to calibrate photometry to the standard system. In the interest
of readability, the discussion of this has been moved to an appendix.
In Table \ref{phottab}, we summarize the calibrated out-of-eclipse
photometry for the eclipsing binary KIC 9777062 from various sources.

\begin{deluxetable*}{lcccc}
\tablewidth{0pt}
\tablecaption{Out-of-Eclipse Photometry for KIC 9777062\label{phottab}}
\tablehead{\colhead{Filter} & \colhead{EHK/2MASS} & \colhead{This Paper} & \colhead{G99} & \colhead{J13}}
\startdata
$U$ & $12.630\pm0.021$ &                  & $12.711\pm0.054$ & $12.680\pm0.006$\\
$B$ & $12.616\pm0.024$ & $12.600\pm0.002$ & $12.598\pm0.017$ & $12.575\pm0.011$\\
$V$ & $12.264\pm0.018$ & $12.241\pm0.002$ & $12.236\pm0.001$ & $12.230\pm0.008$\\
$R_C$ &                &                  & $11.897\pm0.025$ &\\
$I_C$ &                & $11.830\pm0.003$ &                  & $11.833\pm0.005$ \\
$J$ & $11.552\pm0.022$ & & &\\
$H$ & $11.448\pm0.018$ & & &\\
$K_S$ & $11.388\pm0.018$ & & &\\
\hline
$V_A$ & & $12.74$ & $12.73$ & $12.73$ \\
$(B-V)_A$ & & $0.33$ & $0.33$ & $0.33$ \\
$(V-I_C)_A$ & & $0.37$ &  & $0.35$ \\
$V_B$ & & $13.33$ & $13.32$ & $13.32$ \\
$(B-V)_B$ & & $0.41$ & $0.41$ & $0.40$ \\
$(V-I_C)_B$ & & $0.49$ &  & $0.47$
\enddata
\tablecomments{EHK: \citet{EHK}. 2MASS: \citet{2mass}. G99: \citet{glush}. J13: \citet{janes}.}
\end{deluxetable*}

\subsection{Spectroscopy of KIC 9777062}

Spectroscopic observations were obtained at the Hobby-Eberly Telescope
(HET) with the High Resolution Spectrograph (HRS; \citealt{tull}) as
part of normal queue scheduled observing \citep{shetet}, the Nordic
Optical Telescope (NOT) Fibre-fed Echelle Spectrograph (FIES;
\citealt{fies}), and the MMT Hectoechelle \citep{hect}.

The configuration of the HET HRS was chosen based upon the spectral
line widths and strength of the secondary in the first spectrum taken.
KIC 9777062 was observed with the configuration 
HRS\_60k\_central\_600g5822\_2as\_2sky\_IS0\_GC0\_2x1 to achieve
resolution $R=60000$.  This configuration covers 4825 \AA\ to 6750
\AA\ with a small break at 5800 \AA\ between the red and blue
CCDs. Exposure times were 540 or 580 seconds.
The data were reduced using the echelle package within IRAF for
standard bias and scattered light removal, 1-D spectrum extraction,
and wavelength calibration.  A nightly correction was made based on
observations of radial velocity standard stars, although this was
a small correction (typically between 0.2 and 0.4 km s$^{-1}$).

FIES covers the spectral range 3700 to 7300 \AA, and we used the
high-resolution mode ($R = 67000$).  Each observing epoch consisted of
two exposures of 1800 s each. The observations were split to reduce
problems with cosmic rays hitting the CCD.  The science exposures were
preceded by a Th-Ar exposure for accurate wavelength
calibration.  In daytime before or after each observing night
calibration images (7 bias and 21 halogen flats) were obtained.  The
FIES data were reduced using the FIEStool pipeline\footnote{\tt
  http://www.not.iac.es/instruments/fies/fiestool/FIEStool.html},
which is a program written in Python to perform bias subtraction, flat
fielding, and scattered light subtraction, and using IRAF tasks (via
PyRAF) from the echelle package to do spectral order tracing,
extraction, and wavelength calibration.

The observations with the Hectochelle multi-object spectrograph
covered 5150 to 5300 \AA at a resolution of 45000. The science
exposures were bracketed by Th-Ar lamp exposures to optimize the
wavelength calibration, and preceded by a dome-lamp flat exposure used
for tracing the spectra on the CCD.

The radial velocities were derived using a spectral disentangling code
written in Python following the algorithm described in \citet{gonz}.
In each iteration step the spectrum for each component is isolated by
aligning the observed spectra using the measured radial velocities for
that component and then averaging. After the first determination of
the primary star spectrum, the contribution from one component can be
subtracted during the determination of the average spectrum of the
other. The radial velocities can also be remeasured using the spectra
with one component subtracted. This procedure is repeated for both
components and continued until a convergence criterion is met.  We use
the broadening function formalism \citep{bfs} to
measure the radial velocities using synthetic spectra from the grid of
\citet{coelho} as templates.

The FIES spectra are mostly taken in pairs at 20 different orbital
phases sampling the orbit and velocity range well. Both the HRS and
the Hectoechelle spectra are also well distributed in phase, but with
fewer observations. The FIES velocities were calculated as the robust
mean measured in 35 spectral orders (the orders 20-55 except order 39
where the wide H$\beta$ line is). We use these 35 orders out of 79
because the signal-to-noise ratio is too low in the bluest orders and
in the reddest orders the spectra are contaminated by telluric
lines. For the uncertainty estimates we use the standard error of the
mean (standard deviation divided by the square root of the number of
measurements) calculated using all 35 orders.  Final FIES velocities
were recalculated during three iterations of the whole disentangling
procedure in which the uncertainties of the primary velocities were
used as weights when averaging the component spectra. This is
reasonable because we have a large set of spectra with similar
signal-to-noise (and hence a small spread in RV uncertainties).
We present the velocities with their measurement uncertainties in Table \ref{spectab}.
The means of the velocity uncertainties are $168\pm34$ m s$^{-1}$ and 
$290\pm140$ m s$^{-1}$ for the primary and secondary, respectively.

\begin{deluxetable*}{rrcrc|rrcrc}
\tablewidth{0pt}
\tabletypesize{\scriptsize}
\tablecaption{Radial Velocity Measurements}
\tablehead{\colhead{mJD\tablenotemark{a}} & \colhead{$v_A$} & \colhead{$\sigma_{A}$} & \colhead{$v_B$} & \colhead{$\sigma_B$} & \colhead{mJD\tablenotemark{a}} & \colhead{$v_A$} & \colhead{$\sigma_{A}$} & \colhead{$v_B$} & \colhead{$\sigma_B$}\\
& \multicolumn{4}{c}{(\kms)} & & \multicolumn{4}{c}{(\kms)}}
\startdata
\multicolumn{5}{c}{NOT FIES Observations} &       6518.521145 &   51.66  & 0.15 & $-43.50$ & 0.52\\ 
6080.615695 &   28.24  & 0.15 & $-16.96$ & 0.20 & 6518.551091 &   51.91  & 0.14 & $-43.62$ & 0.30\\ 
6086.702696 & $-32.85$ & 0.12 &   52.73  & 0.17 & 6520.462688 &   72.09  & 0.15 & $-66.35$ & 0.23\\ 
6092.678310 &   18.32  & 0.19 &  $-4.24$ & 0.34 & 6520.484531 &   72.09  & 0.14 & $-66.56$ & 0.28\\ 
6093.628552 &   28.93  & 0.23 & $-15.01$ & 0.33 & 6521.526209 &   69.33  & 0.19 & $-65.40$ & 0.33\\ 
6109.611515 &  $-4.47$ & 0.23 &   23.06  & 0.42 & 6521.548053 &   68.88  & 0.25 & $-63.26$ & 1.04\\ 
6115.580028 &   62.44  & 0.16 & $-54.49$ & 0.31 & 6543.464550 & $-18.18$ & 0.17 &   35.86  & 0.21\\ 
6115.602482 &   62.78  & 0.18 & $-54.28$ & 0.42 & 6543.486397 & $-18.89$ & 0.17 &   36.34  & 0.22\\ 
6123.583515 & $-40.60$ & 0.19 &   61.51  & 0.28 & 6549.378082 & $-25.21$ & 0.15 &   44.76  & 0.18\\ 
6123.607625 & $-40.55$ & 0.24 &   62.07  & 0.42 & 6549.399931 & $-24.95$ & 0.15 &   44.43  & 0.21\\ 
6227.334731 &   19.16  & 0.17 &  $-6.34$ & 0.21 &   \multicolumn{5}{c}{HET HRS Observations} \\
6389.680638 & $-19.69$ & 0.17 &   38.22  & 0.22 & 6038.914740 &   65.17  & 0.15 & $-58.16$ & 0.50\\
6389.702489 & $-20.05$ & 0.15 &   38.76  & 0.23 & 6039.908239 &   73.04  & 0.18 & $-66.09$ & 0.28\\
6396.682755 & $-16.22$ & 0.20 &   35.42  & 0.38 & 6078.813734 &   72.98  & 0.15 & $-67.87$ & 0.29\\
6396.704618 & $-16.18$ & 0.15 &   35.53  & 0.35 & 6084.790102 & $-41.30$ & 0.14 &   60.92  & 0.34\\
6425.623269 &   64.88  & 0.20 & $-57.38$ & 0.32 & 6109.954081 &  $-1.20$ & 0.16 &   16.74  & 0.27\\
6425.645114 &   64.44  & 0.27 & $-56.79$ & 0.29 & 6110.719643 &    7.09  & 0.24 &    7.41  & 0.22\\
6428.662997 & $-30.15$ & 0.13 &   49.82  & 0.22 & 6116.719453 &   72.61  & 0.19 & $-64.99$ & 0.29\\  
6428.684842 & $-30.39$ & 0.13 &   50.32  & 0.26 & 6123.917979 & $-39.15$ & 0.16 &   59.74  & 0.25\\  
6429.666715 & $-40.33$ & 0.14 &   62.74  & 0.33 & 6125.682043 & $-29.63$ & 0.15 &   48.44  & 0.25\\  
6429.688566 & $-40.29$ & 0.15 &   63.09  & 0.18 & 6126.683770 & $-22.70$ & 0.16 &   39.95  & 0.21\\  
6430.621638 & $-41.99$ & 0.15 &   64.57  & 0.22 & 6131.655552 &   24.09  & 0.14 & $-11.02$ & 0.17\\  
6430.643464 & $-41.91$ & 0.16 &   64.58  & 0.25 & 6133.878550 &   51.13  & 0.45 & $-41.68$ & 0.56\\  
6445.675337 &   40.42  & 0.15 & $-28.68$ & 0.15 & \multicolumn{5}{c}{MMT Hectoechelle Observations}\\
6445.697184 &   39.51  & 0.14 & $-27.96$ & 0.16 & 4401.617194 &   31.90  & 0.75 & $-20.94$ & 0.82 \\ 
6457.644779 &   13.90  & 0.16 &  $-1.35$ & 0.24 & 5374.659578 & $-36.13$ & 0.65 & 56.06 & 2.03 \\    
6457.666628 &   14.11  & 0.14 &  $-1.25$ & 0.18 & 5457.741551 &   13.62  & 1.07 & 2.10 & 1.43\\      
6489.691312 & $-37.51$ & 0.17 &   58.16  & 0.18 & 5812.757900 & $-19.52$ & 1.11 & 37.18 & 1.19\\     
6489.711426 & $-37.64$ & 0.16 &   57.72  & 0.24 & 5812.827731 & $-21.18$ & 0.87 & 39.15 & 1.41\\     
6507.524857 & $-41.91$ & 0.17 &   64.16  & 0.27 & 6022.842012 &   31.24  & 0.79 & $-18.81$ & 1.29\\  
6507.546702 & $-41.83$ & 0.17 &   64.00  & 0.26 & 6023.883682 &  $-6.91$ & 0.71 & 25.53 & 1.25\\     
6518.448366 &   50.69  & 0.13 & $-42.73$ & 0.31 & 6024.954796 & $-31.89$ & 0.77 & 54.52 & 1.28
\enddata
\label{spectab}
\tablenotetext{a}{mJD = BJD - 2450000.}  
\end{deluxetable*}

To calculate velocities from the HRS observations, the spectra were split
into nine segments and the same procedure was followed with the
exception that we did not apply the velocity uncertainties as
weights. All were given equal weight with two exceptions given zero
weight: one because of very low signal-to-noise and continuum
normalization difficulties, and the other because of very low velocity
separation between the two components. The means of the uncertainties
are $181\pm84$ m s$^{-1}$ for the primary and $300\pm110$ m s$^{-1}$
for the secondary.

Although the Hectoechelle spectra from the MMT have lower
signal-to-noise than the other spectra, we were able to extract radial
velocities for both stars at each epoch.  Velocities were calculated
in the same way as the HRS spectra, but split into seven
segments. The uncertainty estimate stated for these measurements is
just the rms of the mean value from the sub-sections except in three
cases with clear influence of an outlier for which a robust standard
deviation was used instead. The means of the uncertainties are
$840\pm150$ m s$^{-1}$ and $1330\pm315$ m s$^{-1}$ for the primary and
secondary, respectively.

\begin{figure*}
\includegraphics[scale=0.5]{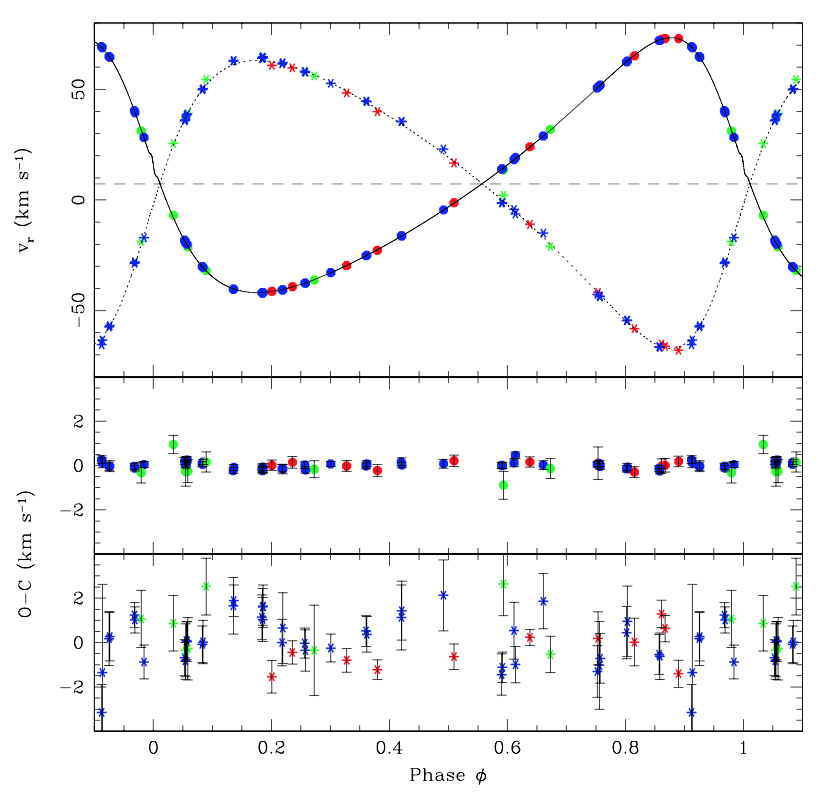}
\caption{Phased radial velocities for KIC 9777062, along with the
  best fit model. Observations of the primary star use dots, and
  secondary star observations use asterisks. Red, blue, and green colors
  are observations from the HET, NOT, and MMT, respectively. The lower
  panels show the observed minus computed values with error bars
  scaled to give a reduced $\chi^2=1$ (see \S
  \ref{binary}).\label{rvplot}}
\end{figure*}

The phased radial velocity measurements are plotted in
Fig. \ref{rvplot}, showing the eccentricity of the orbit.  In this
situation where we are making use of radial velocities from three
different telescopes and instrument configurations, systematic errors
can be as important as random ones. We therefore looked at zeropoint
offsets between a best fit orbit model and different subsets of
data. For measurements from the NOT (both primary and secondary star)
and primary star measurements from the HET and Hectoechelle, we find
mean offsets of less than 0.1 km s$^{-1}$. For measurements of the
secondary star from HET, there was a larger shift of $-0.57$ km
s$^{-1}$, and $+0.52$ km s$^{-1}$ from the Hectoechelle. We opted not
to correct for these offsets because the more precise primary star
measurements agree better with the NOT data and the model.

\section{Analysis of KIC 9777062}\label{analy}
\subsection{Cluster Membership}

For a cluster like NGC 6811 where the stars are spread over a large
area, it is especially important to pay close attention to cluster
membership criteria.  \citet{janes} presented photometric and
structural data on the cluster. This binary star is projected {\bf
  $13\farcm4$} from their cluster center ($\alpha_{2000} = 19^{\rm
  h}37^{\rm m}17^{\rm s}$,
$\delta_{2000}=+46\degr23\arcmin18\arcsec$), which places it well
outside the cluster core, about twice the effective radius determined
from exponential or Plummer model fits \citep{janes}. Because it lies
in the outskirts of the cluster, it is important to examine other
indicators of cluster membership.

The proper motion of KIC 9777062 was measured by \citet[][star
  195]{sanders6811}, \citet{dias14}, and \citet{khar13}, finding
membership probabilities of 93\%, 96\%, and 30\%. As for radial velocities,
four studies have determined means for the cluster.  \citet{frinch}
spectroscopically identified 7 likely cluster members and determined a
mean velocity of $+6.03\pm0.30$ \kms (where the error of the mean is
quoted), \citet{merm08} found $+7.28\pm0.11$ \kms from three stars,
\citet{mz} found $+6.68\pm0.08$ km s$^{-1}$, and \citet{meibom13} find
$+7.7\pm0.04$ km s$^{-1}$ from hundreds of stars.
In addition, 6 giants we identify as likely
cluster members have APOGEE radial velocities in Data Release 12 that
give a mean value of $+7.70\pm0.30$ km s$^{-1}$. KIC 9777062 was
actually observed by \citet{frinch} (identified as TYC 3556-00370-1),
who found a velocity of $+5.92\pm1.82$ km s$^{-1}$. Based on the date
of their observation (UT 2003 Sept. 16), this was shortly before
secondary eclipse (phase $\phi=0.51$), so their measurement was
probably representative of the system velocity of the binary. The
system velocities $\gamma$ for the components of the binary were
fitted separately (see \S \ref{binary}), and the values ($+7.2$ and
$+7.7$ km s$^{-1}$) are fully consistent with the \citeauthor{merm}
and \citeauthor{meibom13} mean velocities, but less so with the
\citeauthor{frinch} and \citeauthor{mz} values. One of our HET spectra
was also taken quite close to a velocity crossing, and also indicates
a velocity of around $+7.2$ km s$^{-1}$. Due to the difficulty of
deriving a good cluster mean and velocity dispersion in the face of a
large spectroscopic binary population, we regard the binary as a
radial velocity member of the cluster. To summarize, the kinematic
evidence points to likely cluster membership.

\subsection{The Ephemeris And Search for a Third Body}

We examined the eclipse timings from the \kep data and ground-based
radial velocities in order to look for any evidence of perturbations
to the ephemeris due to a third star in a longer orbit about the
eclipsing binary or apsidal motion. We used the method of \citet{kwee}
to determine the times of mid-eclipse, using short cadence photometry
where available to achieve greater timing precision and long cadence
photometry in all other cases. The eclipse timings are given in Table
\ref{etimes}. All eclipse times were put on barycentric julian date
(BJD) barycentric dynamical time (TDB) system,
with ground-based times converted from heliocentric to barycentric
time using an online tool\footnote{\tt
http://astroutils.astronomy.ohio-state.edu/time/hjd2bjd.html}
\citep{bjdcalc}.  Based on the higher precision primary eclipse
timings, there are no signs of deviations from a linear ephemeris that
could indicate the effects of a third body. That best fit ephemeris is
\[ \mbox(Min I) = 2454965.58156(5) + 19.2300409(11) \times E \]
The timing of secondary eclipses is less precise due to their
shallower depths, but the secondary eclipse timings mostly agree with
the primary eclipse timings given a phase $\phi=0.59955$ for the
center of the secondary eclipse. There is significant period
difference though, which may be evidence of apsidal motion:
\[ \mbox(Min II) = 2454977.11083(17) + 19.2300245(41) \times E \]

\begin{deluxetable}{llcl}
\tablewidth{0pt}
\tablecaption{\kep Eclipse Timing Observations}
\tablehead{\colhead{mJD\tablenotemark{a}} & \colhead{$\sigma$ (d)} & \colhead{Eclipse Type} & \colhead{Note}}
\startdata
4965.58156 & 0.00012 & P & Q1 start\\
4977.11100 & 0.00013 & S &     \\
4984.81147 & 0.00029 & P &     \\
4996.34134 & 0.00021 & S &     \\
5004.04140 & 0.00006 & P & Q2 start\\
5023.27175 & 0.00007 & P &     \\
5034.80046 & 0.00020 & S &     \\
5042.50174 & 0.00010 & P &     \\
5054.03076 & 0.00037 & S &     \\
5061.73156 & 0.00011 & P &     \\
5073.26098 & 0.00013 & S &     \\
5080.96170 & 0.00022 & P &     \\
5100.19178 & 0.00008 & P & Q3 start\\
5111.72165 & 0.00023 & S &     \\
5119.42167 & 0.00008 & P &     \\
5130.94986 & 0.00049 & S &     \\
5138.65199 & 0.00012 & P &     \\
5150.18084 & 0.00021 & S &     \\
5157.88190 & 0.00011 & P &     
\enddata
\tablenotetext{a}{mJD = BJD - 2450000.}  
\tablecomments{A portion of this table is published here to demonstrate its 
form and content. Machine-readable versions of the full table are available.}
\label{etimes}
\end{deluxetable}

\begin{figure}
\includegraphics[scale=0.3]{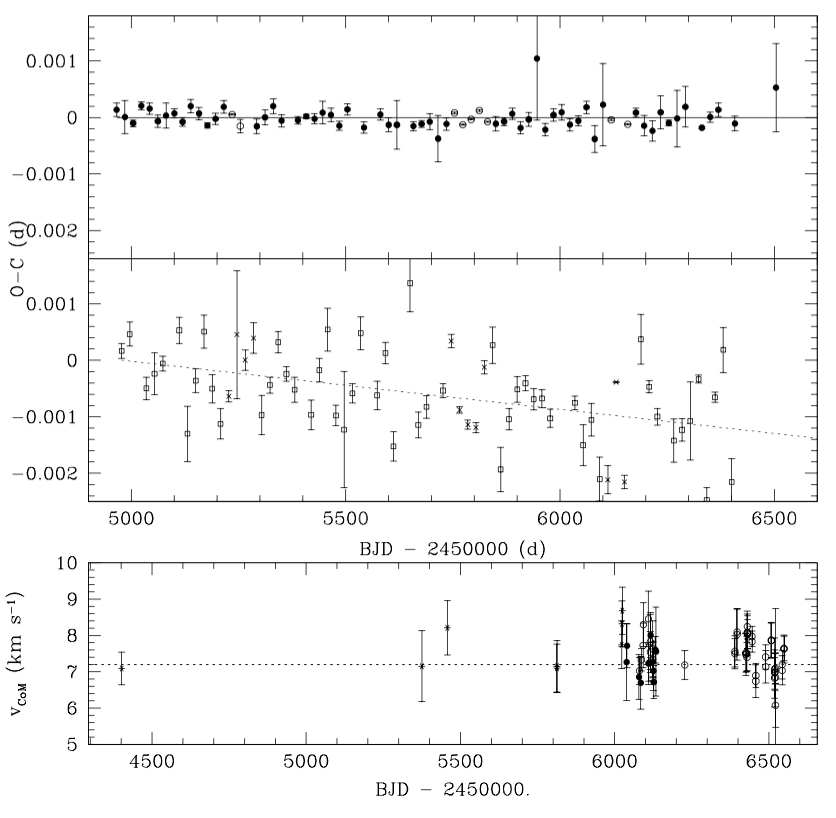}
\caption{{\it Top panels:} Observed eclipse times minus computed
  predictions from the linear best-fit ephemeris for the primary
  eclipses. Primary eclipses are shown with $\bullet$ (long cadence)
  and $\bigcirc$ (short cadence), and secondary eclipses are shown
  with $\sq$ (long cadence) and $\times$ (short cadence). {\it Bottom
    panel:} Center-of-mass radial velocities for the the binary,
  computed using the best-fit mass ratio $q=0.885$. HET observations
  are shown with $\bullet$, NOT observations are shown with
  $\bigcirc$, and MMT observations with $\ast$, with the fitted system
  velocity for the primary shown with a dashed line.
\label{tert}}
\end{figure}

The radial velocity of the binary center of mass provides an
additional means of constraining the presence of an orbiting
companion. Using the best-fit mass ratio from the binary star models
($q = 0.885$), the center-of-mass velocity can be calculated for each
epoch provided the radial velocities of both stars are reliably
measured. The results are shown in Fig. \ref{tert}, but there is as yet no
evidence of systematic shifts produced by a tertiary star.

\subsection{Spectroscopic Information on the Binary}\label{constrain}

Spectroscopic constraints on the temperatures and compositions of the
binary star components are important to modeling of binary
stars (especially for the limb darkening applied to the light curves)
and to the translation of the stellar characteristics into distance and age.
We analysed the disentangled FIES spectra in order to determine the
metallicity and temperature of both stars in the binary.

Spectral disentangling cannot determine the light ratio of
unresolved stars, however, so we employed light ratios in $BVI$
filters from the binary analysis (see Table \ref{chartab}) in order to
properly gauge line depths.  We analysed the spectra of the primary
and secondary stars with the preferred value of the light ratios, and
for ratios one standard deviation higher and lower. In this way we
could estimate the impact of the light ratio on the determined
atmospheric temperatures and surface metallicity. Thus, we analysed
six spectra in total.

These spectra were analysed using the VWA software package
\citep{bruntt2010a, vwa, bruntt2012}, and we briefly summarize the
analysis here. First, a synthetic spectrum is calculated with
preliminary values for $\teff$, $\logg$, and metallicity. We
interpolated the atmospheric model in the MARCS grid \citep{MARCS},
which uses \citet{grevesse} solar abundances, and atomic data were
extracted from the VALD database \citep{vald2015}. The computed
spectrum was rotationally broadened to fit the width of the observed
lines. For the primary we obtained $20\pm1$ km s$^{-1}$, and for the
secondary, $19\pm1$ km s$^{-1}$. This spectrum was compared to the
observed spectrum in order to identify approximate continuum
windows. In this way we renormalized the output spectrum from the
disentangling software using the ``rainbow'' procedure described in
\citet{bruntt2010a}. We note that the renormalization was repeated
after improving the atmospheric parameters.

Second, we used VWA to automatically select isolated lines in the
spectrum, and each line was fitted to determine the abundance. For
each model fit we keep $\teff$, microturbulent velocity, and metallicity
fixed for each run. It is important to note that we did not change
$\logg$; instead we used the well-determined values from the binary
star analysis ($\logg = 4.16$ and 4.22 for the primary and secondary,
respectively). After running several models with different
$\teff$\ and microturbulence, we then used the neutral and singly-ionized
iron lines (Fe I and Fe II) to infer the adjustment needed to
eliminate any correlation between Fe I abundance and either excitation
potential or equivalent width. In addition, we
required that Fe II and Fe I lines yield the same mean value for the
abundance. In the case of Fe I lines we used a slight adjustment of
the abundance due to NLTE effects, following the calculations of
\cite{nlte1996}. Interpolating in the figures from \cite{nlte1996}, we
added 0.07 dex to the Fe I abundances from VWA.

The atmospheric parameters we determined are given in Table
\ref{abunds}, including the uncertainties due to the light ratio and
scatter in the Fe I and Fe II abundances.  Our analysis found a fairly
standard value for the microturbulent velocity $\xi$ for the secondary
star ($2.0\pm0.2$ km s$^{-1}$), but we found it necessary to use a
larger value of $3.7\pm0.2$ \kms for the primary star. This may be
related to possible convective blueshifts in the radial
velocities of the primary star, as discussed in \S \ref{binary}.

\begin{deluxetable}{lcc}
  \tablecolumns{3}
  \tablewidth{0pc}
  \tablecaption{Spectroscopic Parameters and Abundances for Eclipsing Binary Components}
  \tablehead{\colhead{} & \colhead{Primary} & \colhead{Secondary}}
\startdata
$\logg$ (adopted) & 4.16 & 4.22\\
$\teff$ (K) & $7700\pm150$ & $7150\pm100$\\
$v_{rot} \sin i$ (km s$^{-1}$) & $20\pm1$ & $19\pm1$\\
$\xi$ (km s$^{-1}$) & $3.7\pm0.3$ & $2.0\pm0.2$ \\
$\mbox{[Fe/H]}$ & $+0.46\pm0.13$ & $-0.03\pm0.13$\\
$\mbox{[Ca/H]}$ & $-0.80\pm0.15$ & $-0.14\pm0.15$\\
$\mbox{[Ti/H]}$ & $-0.10\pm0.16$ & $-0.03\pm0.15$\\
$\mbox{[Cr/H]}$ & $+0.35\pm0.14$ & $-0.20\pm0.15$\\
$\mbox{[Ni/H]}$ & $+0.71\pm0.13$ & $-0.27\pm0.14$
\enddata
\label{abunds}
\end{deluxetable}

The rms scatter in Fe I line abundances was 0.19 for the primary and
0.21 dex for the secondary. This is a relatively high scatter, and
likely due to an imperfect disentangling of the
spectra. Contributors to the uncertainty on the mean Fe abundance are
0.02 dex due to line-to-line scatter ($\sigma=0.19/\sqrt{86}$), 0.10
dex due to uncertainties in $\teff$\ and microturbulence, and 0.03 due
to the light ratio. Furthermore, we adopted a 0.07 dex systematic
error from \cite{vwa} due to the adopted atmospheric models,
NLTE effects, and continuum normalization errors. We add these
contributions in quadrature and find the estimated uncertainty on the
metallicity to be 0.13 dex for both components. The uncertainty on the
elements Ca, Ti, and Cr listed in Table \ref{abunds} are higher
because only 4-8 lines were used. Our Ni abundance is based on 30 lines for
the primary and 15 for the secondary.

\begin{figure*}
\includegraphics[scale=0.4,angle=90]{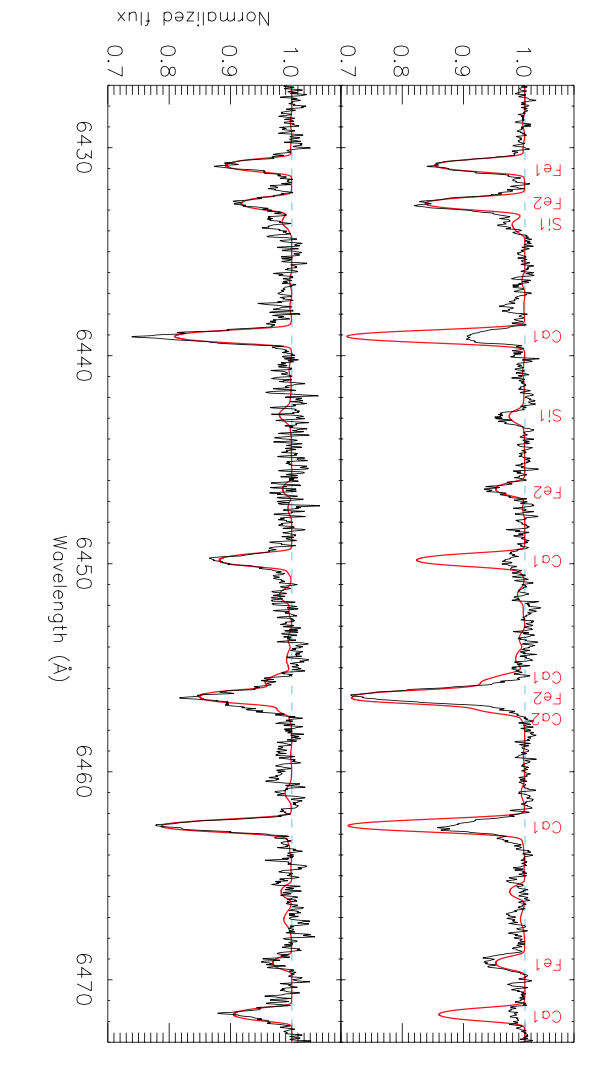}
\caption{Comparison of disentangled spectra for the KIC 9777062
  primary {\it (top panel)} and secondary {\it (bottom panel)} with
  best-fit synthetic spectra ({\it red lines}) in the vicinity of
  four Ca I lines.
\label{disent}}
\end{figure*}

In Figure \ref{disent}, we compare the observed spectra of the two
stars with synthetic spectra computed using the parameters listed in
Table \ref{abunds}. Note that for all elements in the figure's
synthetic spectra, the abundances were scaled with the Fe abundance.
The secondary star appears to have an approximately solar Ca to Fe
abundance ratio, but the primary star is severely underabundant in Ca
relative to Fe. Ni and Cr, on the other hand, are quite overabundant
relative to the Sun.
Given that other spectroscopic studies of NGC 6811 stars
\citep{mz,occam} find abundances close to that of the secondary star,
the primary star appears to be anomalous. The signatures of high metal
abundances and low Ca abundance identify the primary as an metal-lined
Am star. The leading explanation for this phenomenon is diffusion via
gravitational settling and radiative levitation
\citep[e.g.,][]{turcotte98,richer}. Calcium tends to be underabundant because
it has a closed electron shell configuration at the temperature of
interest, reducing its radiative cross section near the peak
wavelengths of the star's emission. Thus, Ca, unlike Fe and other
metallic species, gravitationally settles into deeper layers.

A-type stars are generally rapid rotators, but when they are not,
chemical peculiarities are frequently present: Am stars, for example,
generally have rotational speeds below 100 \kms \citep{abt,astarrot}.
Both stars in the binary rotate at speeds well below this threshold.
That said, both stars are rotating too fast to
have pseudo-synchronized to their orbital angular velocities at
periastron passage (8.2 and 7.1 km s$^{-1}$; \citealt{hut}).
So tidal effects appear to have forced
the rotation rates lower despite the fact that the theoretical predictions
for synchronization by dynamical tides \citep{zahn} give timescales
that are several orders of magnitude longer than the $\sim1$ Gyr
age of the cluster. \citet{carq} came to similar
conclusions for a sample of field Am stars in binaries. The
fact that close to two-thirds of Am stars are members of spectroscopic
binaries \citep{cp} indicates that tidal effects are an important
ingredient in this phenomenon.

For the purposes of our modeling below, this means that the primary
star spectrum will not help constrain the bulk (interior) chemical
abundances of the stars. We are still left with the measurements of
the secondary star and literature measurements of other stars in the
cluster, and these are consistent with each other to about $1\sigma$
of the secondary star measurement error.

The derived spectroscopic temperatures are $7700\pm150$ K and
$7150\pm100$ K for the primary and secondary stars.  We computed
photometric temperature estimates as checks on these values.
Unfortunately, there are fewer well-studied stars for calibrating
A and early F stars than there are for G, K, and M spectral types.  We
employed relations given by \citet{casagrande} and \citet{boyajian},
although it should be noted that there are roughly twice as many A
star calibrators in the \citeauthor{boyajian} paper and thus their
value should be considered to be better constrained.
We used the system photometry, luminosity ratios derived from the
binary modeling, metallicity [Fe/H]$=+0.04$, and an average reddening $E(B-V)=0.07\pm0.02$
(see \S \ref{chemred}) in empirical relations using optical colors.

Using \citet{casagrande} relations, we used the $(B-V)$
color to calculate the primary star (7330 K)
and secondary star (6950 K) temperatures.  These are
$1-2\sigma$ smaller than the spectroscopic values for both stars, but
the stars are at the blue end of the quoted range of reliability for
the color-temperature relations. [The $(V-I_C)$ color could not be
  used because the stars were outside the color range.]
Using \citet{boyajian} relations, temperatures from $(V-I_C)$ colors
(around 7590 and 6990 K for the primary and secondary stars,
respectively) are consistent with the spectroscopic values, but the
six parameter $(B-V)$ color relation produces much lower values (7040
and 6690 K, respectively). There are, however, few calibrators in the
vicinity of the color of our stars, and the fit may be going too
low. In addition, the uncertainty due to the reddening is
approximately 100 K using $B-V$ colors.

The Am phenomenon operating in the primary star is known to produce
line blanketing in the blue part of the spectrum, and this makes an Am
star redder than it would otherwise have been. Using the $(B-V)-\teff$
calibration specifically for Am stars from \citet{netopil}, we derive
$\teff = 7620 \pm 175$ K for the primary star, in excellent agreement
with the spectroscopic temperature.

There are benefits in also comparing the spectroscopic
temperatures with stars very similar to the ones in KIC 9777062 in
well-studied eclipsing binary systems: single stars in this range of
spectral types tend to be rapidly rotating, which can produce
significant gravity darkening and effects due to the angle of
observation.
An excellent comparison is XY Cet B \citep{south11}. This star has
nearly identical mass (0.2\% difference) and similar radius (1.5\%
difference, with XY Cet B larger) as KIC 9777062 A, and both are Am
stars. \citeauthor{south11} found that the characteristics of XY Cet
agree best with isochrones for a solar bulk composition and age
slightly less (850 Myr) than NGC 6811. But because both stars in XY
Cet are Am stars, the bulk composition cannot be directly determined,
which could impact the temperatures and the age determination.
Otherwise, these indicators point to unusually good agreement between
XY Cet B and KIC 9777062 A. \citeauthor{south11} adopt a temperature
of $7620 \pm 125$ K based on comparisons with synthetic spectra and
Str\"{o}mgren photometry, in good agreement with our spectroscopic
temperatures.

KIC 9777062 B is similar (less than 2\% difference in mass and
radius) to HY Vir B, with HY Vir B the lower mass star
\citep{slacy}. The HY Vir system appears to be similar in age to NGC
6811, but probably slightly older (1.35 Gyr). Although the chemical
composition for HY Vir has not been spectroscopically measured, the
authors find that a super-solar metallicity fits the masses and radii
of the stars in the binary.  Lower mass, higher age, and higher
metallicity would all push the star in the direction of lower
temperature, so their quoted value for HY Vir B ($6550 \pm
120$ K) should be considered a lower limit for KIC 9777062 B.
On the other hand, V501 Mon B \citep{torres15} is approximately 2\%
larger in mass and has a very similar age (1.1 Gyr) and composition
([Fe/H]=+0.01). Its temperature ($7000\pm90$ K) is roughly consistent
with that of KIC 9777062 B.

In conclusion, we are choosing to depend on the spectroscopically
derived temperatures in our analysis below. This decision is partly
based on the difficulties in using photometry to determine
temperatures for A stars: there are currently small numbers of
calibrators for A stars, disagreement between color-temperature
relations from different researchers, and questions of how applicable
single star calibrations can be for slowly rotating stars (as in our
eclipsing binary). However, the photometric temperature for the
  primary from an Am star calibration does support our spectroscopic value.

\subsection{Binary Star Modeling}\label{binary}

To simultaneously model the ground-based radial velocities and photometry and
\kep photometry, we used the ELC code \citep{elc}. 
We fitted the binary with a primary set of 14 parameters: orbital
period $P$, reference time of conjunction (primary eclipse) $t_c$,
velocity semi-amplitude of the primary star $K_1$, mass ratio $q = M_2
/ M_1 = K_1 / K_2$, systematic velocities for both stars $\gamma_1$
and $\gamma_2$, eccentricity $e$, phase difference between eclipses
$\Delta \phi$, inclination $i$, ratio of the primary radius to average
orbital separation $R_1 / a$, ratio of radii $R_1 / R_2$, temperature
ratio $T_2 / T_1$, and contamination of the \kep light curve in the
two seasons of short cadence observations.  The \kep contamination
parameters are used to account for possible dilution of the light
curves by stars that are not physically associated with the
binary. \kep pixels and stellar point-spread functions are large by
the standards of most ground-based cameras, and it is more difficult
to disentangle the light of blended stars. The contamination
parameters did not play a major role in the fits, however, because
there are no bright contaminating stars nearby. In the fits, the
contamination parameters themselves are found to be quite small (0.03\%
and 0.45\% for the two seasons), in accord with values from the \kep
Input Catalog (see \S \ref{kphot}).

\begin{figure}
  \includegraphics[scale=0.3]{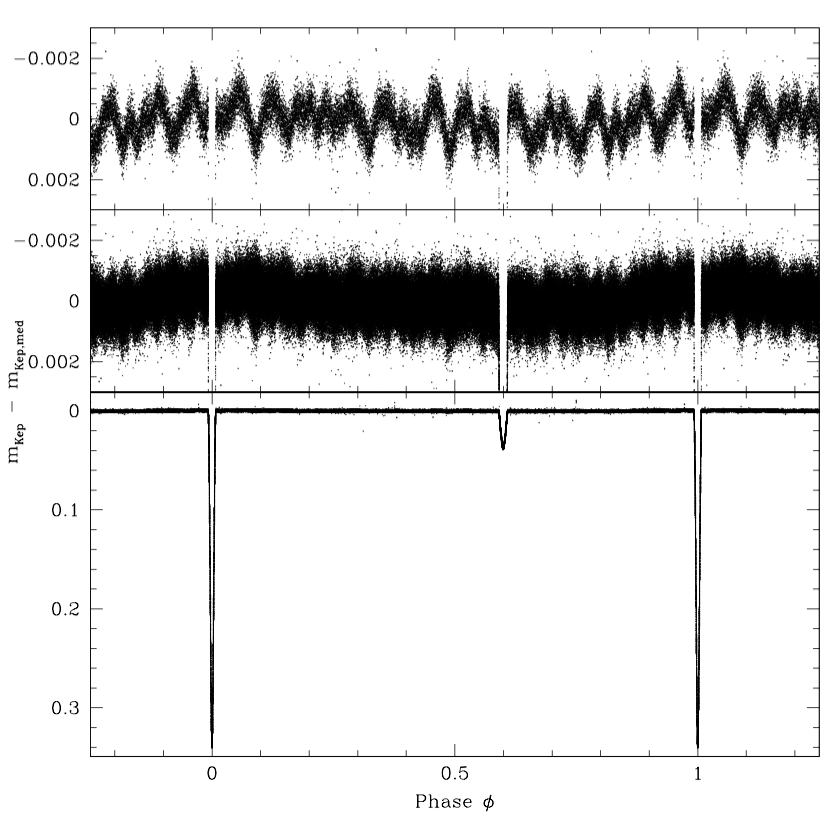}
\caption{Short cadence \kep data versus orbital phase. {\it Top panel:} One orbit cycle from quarter 10, showing pulsations. {\it Middle panel:} All data, zoomed on the out-of-eclipse variations, showing the reflection effect. {\it Bottom panel:} All data, showing the positions and depths of eclipses.
\label{reflecteffect}}
\end{figure}

The light curve is quite flat outside of eclipses, with a small
reflection effect (full amplitude about 0.001 mag) observable in the
\kep data (see Fig. \ref{reflecteffect}). The maximum occurs around
primary eclipse, with a minimum centered on the secondary eclipse. The
variation is asymmetic, in accord with the eccentricity of the
orbit. While this type of variation can provide critical information
for binaries that do not eclipse \citep{BEER}, its physical content is
outweighed by the eclipse observations, and we removed the small
reflection effect from our light curves and restricted our modeling to
eclipses and phases shortly before and after them.

Gravity darkening should have a negligible effect on the light curves
due to the slowly rotating, nearly spherical stars in this fairly wide binary.
However, limb darkening plays an important role in the determination
of the system parameters because only the limb of the secondary star
is eclipsed, thanks to the eccentricity of the orbit and its
orientation with respect to the line of sight. In turn, the
shallowness of the secondary eclipse results in a relatively low
signal-to-noise ratio in measurements. As a result, we find that we
are unable to reliably fit for the coefficients of a limb darkening
law in any passband except {\it Kepler}'s. The most important effect
of this is on the determination of the luminosity ratios that are
needed to constrain the color-magnitude positions of the two stars (\S
\ref{cmd}).

To assess the possible systematic errors, we approached the limb
darkening in several ways.  A common problem in light curve fitting is
that the coefficients of the limb darkening laws are correlated, and
fits can sometimes even leave the regime of realistic values. If
theoretical values are used, systematic errors in $\teff$, log $g$, or
composition will lead to systematic errors in fitted parameters. Given
the chemical peculiarities seen for the primary star, it is reasonable
to wonder whether we could rely on any theoretical coefficients even
if $\teff$ and log $g$ were perfectly known.  In our first approach
(inspired by \citealt{brogaard}), we used a quadratic limb darkening
law for each passband, held the $BVI_C$ coefficients fixed at values
determined from ATLAS atmospheres \citep{claret} for the estimated
effective temperature and gravity of the star, and fitted for one
parameter for each star in the \kep passband. Even if our fixed
\kep coefficients are in error, this should be largely
compensated for by a change in the other (correlated) parameter. In
the second approach, we again used a quadratic law, but used the
algorithm of \citet{kipping} to fit for both coefficients for both
stars in the \kep passband.  \citeauthor{kipping} recasted the
limb-darkening coefficients in a way that reduces correlation, and
which allows for straightforward sampling of the triangular region of
physically realistic values. This last point means that our ignorance
about the precise form of the limb darkening for the stars can be fully
accounted for in the uncertainties of the derived parameters of the
binary system. Kipping's method has been implemented in the ELC code.
In the third method, PHOENIX atmosphere models \citep{hauschildt} were
used to calculate the relative contributions from the two stars to the
total flux, but the limb darkening coefficients were fitted using the
\citet{mandel} algorithm. This approach is somewhat inconsistent in
that the limb darkening in the atmosphere models will not be the same
as used in fitting the light curve. However, the relationship between
the fitting parameters and the relative contributions to the total
system luminosity will be based on a more sophisticated description of
the stellar atmospheres.

We applied the spectroscopic temperature of the primary star $T_1$
(see \S \ref{constrain}) as an observed constraint, allowing the
value to vary, but applying a $\chi^2$ penalty if the value deviated
from the input.
From numerical experiments, we found that the models were generally
pushed toward temperatures at the high end of the range ($7800 - 7900$
K) by the combination of \kep photometry and ground-based observations
of the primary eclipse in at least one filter band, despite the
$\chi^2$ penalty. The depth of the ground-based primary eclipse
observations helps set the temperature in combination with the
temperature ratio information from the relative sizes of the primary
and secondary eclipses derived from the precise \kep
observations. These shifts are within about $1\sigma$ of the spectroscopic
temperature, however.

The quality of the model fit was quantified by an
overall $\chi^2$, and the minimum value was sought first using a
genetic algorithm \citep{metcalfe99,gene,geneelc} to explore a large
swath of parameter space, followed by Markov chain Monte Carlo
modeling \citep{mcmc} to explore models near the minimum and
to estimate the uncertainties in the binary model parameters. The
quoted parameter uncertainties are based on the range of values that
produce a total $\chi^2$ within 1 of the minimum value, which
approximates a $1\sigma$ uncertainty \citep{avni}.  The same procedure
was used to determine uncertainties in the masses and radii of both
stars.

Because the uncertainties for the measurements are used in the
calculation of $\chi^2$, it is important that these uncertainties are as
realistic as possible. We therefore scaled the uncertainty estimates
for the radial velocity and photometric observations in order to
return a reduced $\chi^2$ value of 1 for each type of
measurement. (For example, the reduced $\chi^2$ value for \kep
photometry was used to scale the \kep photometry uncertainties,
the reduced $\chi^2$ value for the NOT radial velocities of the
primary star was used to scale those uncertainties, and so on.) This
procedure forces the measurement uncertainties to be consistent with
the observed scatter around a preliminary best fit model. After this
scaling (by the square root of the reduced $\chi^2$), we computed
final models to determine the binary model parameters and
uncertainties. 

For the different photometric datasets the scaling factors were
similar: from 1.30 for $B$ up to 1.49 for {\it Kepler}.  This
procedure was especially important for the velocities because the
scatter of the secondary radial velocities around best fit models was
always found to be considerably higher than implied by the measurement
uncertainties. For the NOT velocities, we needed to scale by a factor
of 3.8, and for the HET data, by a factor of 2.1. (For comparison, we
found that the rms variation in the primary star velocities was almost
exactly what was expected from the intrinsic measurement uncertainties
according to a $\chi^2$ test, and the scaling factors were less than
10\%.)  So, there appears to be a source of velocity variability
intrinsic to the secondary star producing this. The variability may be
related to the pulsation of the secondary star (see \S \ref{to}), but
we did not find a correlation between the velocity residuals and the
pulsations in the \kep light curve.  From test models, we find that
the best fit orbit can be pulled from the better determined solution
implied by the primary star. That said, we do not want to discount the
secondary star velocities (and indeed, we cannot if we want to
determine masses for the component stars). To address this, we
independently fitted an orbit to the secondary star velocities and
used the scatter around this solution to scale the velocity
uncertainties (and effectively, the weights) we used in our final
combined fit. This way the secondary star velocities contribute to
determination of parameters like the eccentricity $e$ and the argument
of periastron $\omega$, but we acknowledge that there are non-orbital
effects on them. The reader should be aware that the differences in
weighting the velocities decreased the final masses of about $0.016$
and $0.008 \msun$ for the two stars relative to weights based on
scatter around the combined fit. Because this shift is larger than the
statistical uncertainties in the model fits, we quote
these systematic uncertainties along with our best fit masses.
We also include a systematic contribution to the uncertainty due to
the possible shift in secondary eclipse phase with time (Figure
\ref{tert}): by not modeling this shift we may be introducing scatter
in the \kep eclipse light curves that would appear as a small increase
in the derived eclipse widths and radii. Based on the potential
shifts, we estimate this effect to come to about $0.002\rsun$ for both
stars.

Figures \ref{phot} and \ref{rvplot} show comparisons of the \kep light
curves and radial velocities with the best fit model, and Table
\ref{chartab} shows the parameters of the best fits. For the runs
using Kipping's algorithm for the limb darkening, we identified three
local minima with $\chi^2$ within 1 of the minimum value found. These
minima appear to have involved values of the primary star limb
darkening coefficients and other parameters (like inclination) that
depend on the limb darkening. When reporting the best fit values in
the table, we give one-sided error bars when the two higher local
minima were both on the same side of the apparent global minimum. The
systematic differences between the results using different limb
darkening algorithms are one indicator of our limits in being able to
reduce parameter uncertainties. For our adopted parameters, we use a
weighted mean of the values from the three methods with a quoted
uncertainty from a quadratic summation of the fit uncertainties (from
the $\chi^2$ analysis) and the scatter in the values from the three
methods.

\begin{deluxetable*}{lcccc}
\tablewidth{0pt}
\tabletypesize{\scriptsize}
\tablecaption{Best-Fit Model Parameters for KIC 9777062}
\tablehead{\colhead{Parameter} & \colhead{Quadratic Law} & \colhead{Kipping Algorithm} & \colhead{Atmospheres/MA02} & \colhead{Adopted}}
\startdata
$T_1$ (K) & \multicolumn{3}{c}{$7700\pm250$ (constraint)} \\%
$x_{Kep,1}$ & 0.2210 (fixed) & $$ & $$\\
$y_{Kep,1}$ & $0.4193^{+0.0043}_{-0.0011}$ & $$ & $$\\%
$x_{Kep,2}$ & 0.2767 (fixed) & $$ & $$\\%
$y_{Kep,2}$ & $0.3587^{+0.0023}_{-0.0050}$ & $$ & $$\\%
$x_{B,1}$ & 0.3603 (fixed) & \multicolumn{2}{c}{0.5259 (fixed)} \\
$y_{B,1}$ & 0.3671 (fixed) & \multicolumn{2}{c}{0.2484 (fixed)} \\
$x_{B,2}$ & 0.4033 (fixed) & \multicolumn{2}{c}{0.5369 (fixed)} \\
$y_{B,2}$ & 0.3294 (fixed) & \multicolumn{2}{c}{0.2752 (fixed)} \\
$x_{V,1}$ & 0.2763 (fixed) & \multicolumn{2}{c}{0.3944 (fixed)} \\
$y_{V,1}$ & 0.3535 (fixed) & \multicolumn{2}{c}{0.2226 (fixed)} \\
$x_{V,2}$ & 0.2979 (fixed) & \multicolumn{2}{c}{0.4037 (fixed)} \\
$y_{V,2}$ & 0.3375 (fixed) & \multicolumn{2}{c}{0.2344 (fixed)} \\
$x_{I,1}$ & 0.1270 (fixed) & \multicolumn{2}{c}{0.2032 (fixed)} \\
$y_{I,1}$ & 0.3262 (fixed) & \multicolumn{2}{c}{0.1429 (fixed)} \\
$x_{I,2}$ & 0.1493 (fixed) & \multicolumn{2}{c}{0.2254 (fixed)} \\
$y_{I,2}$ & 0.3255 (fixed) & \multicolumn{2}{c}{0.1572 (fixed)} \\
\hline
$P$ (d) & 19.2300391 & 19.2300390 & 19.2300392 & 19.2300391\\%
$\sigma_P$ (d) & 0.0000002 & 0.0000003 & $_{-0.0000004}^{+0.0000002}$ & 0.0000003\\
$t_C - 2450000$ & 5234.802024 & 5234.802025 & 5234.80202 & 5234.802025\\%
$\sigma_{t}$ & 0.000007 & 0.000010 & 0.00001 & 0.000011 \\
$\gamma_1$ (\kms) & $7.203\pm0.002$ & $7.193\pm0.002$ & $7.193\pm0.002$ & $7.196\pm0.005$\\%
$\gamma_2$ (\kms) & $7.660\pm0.004$ & $7.655\pm0.004$ & $7.657\pm0.005$ & $7.657\pm0.005$\\%
$q$ & $0.885\pm0.002$ & $0.885\pm0.002$ & $0.885\pm0.002$ & $0.885\pm0.002$\\%
$K_1$ (\kms) & $57.58\pm0.03$ & $57.58\pm0.03$ & $57.58\pm0.02$ & $57.58\pm0.03$\\%
$K_2$ (\kms) & $65.06\pm0.12$ & $65.05\pm0.12$ & $65.05\pm0.13$ & $65.05\pm0.13$\\%
$i$ ($\degr$) & $87.2517\pm0.0005$ & $87.2592\pm0.0009$ & $87.255\pm0.001$ & $87.254\pm0.004$\\%
$e$ & $0.3530\pm0.0001$ & $0.3517\pm0.0002$ & $0.3517\pm0.0002$ & $0.3526\pm0.0007$\\%
$\omega$ ($\degr$) & $65.13\pm0.01$ & $65.02\pm0.01$ & $65.02\pm0.02$ & $65.07\pm0.05$\\%
$\Delta\phi$ & 0.599340 & 0.599341 & 0.599342 & 0.599341\\%
$\sigma_{\Delta\phi}$ & 0.000002 & 0.000002 & 0.000003 & 0.000003\\%
$R_1/a$ & $0.04007^{+0.00002}$ & $0.03987_{-0.00001}$ & $0.03990\pm0.00001$ & $0.03991\pm0.00010$\\%
$R_2/a$ & $0.03543_{-0.00002}$ & $0.03534^{+0.00002}$ & $0.03531\pm0.00002$ & $0.03536\pm0.00006$\\%
$R_1/R_2$ & $1.1311\pm0.0008$ & $1.1281\pm0.0008$ & $1.1301\pm0.0002$ & $1.1300\pm0.0015$\\%
$(R_1+R_2)/a$ & $0.07550\pm0.00002$ & $0.07522_{-0.00003}$ & $0.07521\pm0.00002$ & $0.07533\pm0.00015$\\%
$T_2/T_1$ & $0.9237\pm0.0005$ & $0.9223\pm0.0003$ & $0.9233\pm0.0003$ & $0.9229\pm0.0007$\\%
contam. (S0) & $0.00453$ & $0.00424$ & $0.00439$ & \\%
contam. (S2) & $0.00028$ & $0.0$ & $0.00015$ & \\%
\hline
$L_2/L_1(B)$ & $0.5445\pm0.0012$ & $0.5431_{-0.0014}^{+0.0006}$ & $0.5336\pm0.0015$ & $0.540\pm0.006$ \\%
$L_2/L_1(V)$ & $0.5834\pm0.0012$ & $0.5831_{-0.0012}^{+0.0006}$ & $0.5752\pm0.0014$ & $0.581\pm0.005$ \\
$L_2/L_1(I_C)$ & $0.6493\pm0.0013$ & $0.6499_{-0.0011}^{+0.0007}$ & $0.6483\pm0.0012$ & $0.6492\pm0.0008$\\
\hline
$M_1/\msun$ & $1.602\pm0.007$ & $1.604\pm0.006$ & $1.604\pm0.007$ & $1.603\pm0.006\pm0.016$\\%
$M_2/\msun$ & $1.418\pm0.003$ & $1.420\pm0.003$ & $1.420\pm0.003$ & $1.419\pm0.003\pm0.008$\\%
$R_1/\rsun$ & $1.749\pm0.002$ & $1.741\pm0.002$ & $1.742\pm0.002$ & $1.744\pm0.004\pm0.002$\\%
$R_2/\rsun$ & $1.546\pm0.002$ & $1.543\pm0.002$ & $1.542\pm0.002$ & $1.544\pm0.002\pm0.002$\\%
$\log g_1$ (cgs) & $4.157\pm0.002$ & $4.161\pm0.002$ & $4.161\pm0.001$ & $4.160\pm0.003$\\%
$\log g_2$ (cgs) & $4.211\pm0.001$ & $4.213\pm0.002$ & $4.214\pm0.001$ & $4.213\pm0.002$
\enddata
\label{chartab}
\end{deluxetable*}

As described earlier, the decomposition of the system photometry into
the contributions from the individual stars cannot be done particularly
precisely using eclipse depths. We have derived our final luminosity
ratios from a straight mean of the results for the three limb
darkening algorithms, using the dispersion in the results as a measure
of the systematic uncertainty. We find $L_2/L_1(B) = 0.540\pm0.006$,
$L_2/L_1(V) = 0.581\pm0.005$, and $L_2/L_1(I_C) = 0.6492\pm0.0008$.

One of the interesting results of our modeling is the persistent
difference between the system velocities $\gamma_1$ and $\gamma_2$
derived for the two stars, with the primary having a velocity that is
lower by about $0.45-0.55$ km s$^{-1}$. A difference of approximately
this size is present when the velocities from each of the telescopes
are modeled separately, which appears to rule out velocity zeropoint
issues although denser radial velocity sampling of the orbit is needed
to be definitive.  Differences like this can potentially be induced by
gravitational redshifts or convective blueshifts if the properties of
the stars are different enough. Using the binary star analysis, the
difference in gravitational redshifts is only expected to produce a
difference of about 2 m s$^{-1}$, which can be neglected
here. Convective blueshifts result from asymmetries in the motions of
gas near the surface: outward moving parcels are generally hotter,
brighter, and cover a larger fraction of the surface than inward
moving parcels. These motions distort line profiles, but also shift
line centers. Convective blueshifts for the Sun reach 300 m s$^{-1}$
\citep{dravins}, and may reach 1000 m s$^{-1}$ for mid-F stars
according to models \citep{dn90} and spectroscopic observations
\citep{ap}. The microturbulence velocity parameter $\xi$ (an
indication of the local velocity dispersion in the stellar
photosphere) also corroborates the idea that convection is not only
happening in the outermost layers of A stars, but is more vigorous in
Am stars than in cooler stars
\citep{landstreet,landstreet09,gebran}. In fact $\xi$ appears to rise
from around 2 km s$^{-1}$ between 6000 and 7000 K up to a maximum of
about 4 km s$^{-1}$ at $\teff \sim 8000$ K before decreasing to zero
at about 10000 K. Spectral synthesis of KIC 9777062 supports this: the
spectra appear to require $\xi$ of 3.7 and 2.0 km s$^{-1}$ for the
primary and secondary stars. Because the secondary star is a pulsator,
there is a possibility that the pulsation might somehow be causing the
difference in $\gamma$. Based on the nature of the pulsations (see the
top panel of Figure \ref{reflecteffect}), we do not see clear evidence
that it is could be due to a higher probability of taking spectra at
light curve minima. However, denser coverage of the radial velocity
curve should be undertaken to rule out the pulsation.

These behaviors indicate that the KIC 9777062 binary may show
something close to the maximum contrast in convective blueshifts
between two stars. The magnitude of the blueshift difference between
the two stars can reasonably be expected to be on the order of a few
hundred m s$^{-1}$, and it appears to be working in the
theoretically-expected direction. Modeling this is beyond
the scope of this work, but this system demonstrates that well-studied
eclipsing binaries containing Am stars could contribute significantly
to the understanding of the atmospheres of these stars. We have
examined the literature on other eclipsing Am stars, but have not
found as clear a case as this. V501 Mon \citep{torres15} is a system
with parameters similar to KIC 9777062, and would be a good candidate
to further study this effect.

\section{Discussion}\label{disc}

One of the important uncertainties in the discussion below about the
eclipsing binary specifically and the cluster generally is the
metallicity of cluster stars. Previous spectroscopic measurements were
summarized in \S \ref{chemred}.  For the purposes of comparisons with
theoretical models from different sources, we will use heavy element
mass fraction $Z$. This is close to an absolute abundance indicator, and
will therefore minimize systematic differences between isochrones due
to chemical composition. [Fe/H] and [M/H] are relative indicators of
heavy element abundance, of course, and because the assumed solar
abundance values are not the same from isochrone set to set, there
will be systematic differences between models for the same value of
[Fe/H]. There remain unresolved issues in determining the total solar
photospheric heavy element fraction, and this will be a significant
uncertainty in our modeling below. Values range from $Z_\odot =
0.0122$ \citep{asp05} to more recent ones like $Z_\odot=0.0153$
\citep{caffau}.

The APOGEE results for NGC 6811 red clump stars imply $Z \approx
0.0137$ (from the combination of the measured [M/H] and assumed
$Z_\odot=0.0122$), and this will be our preferred choice for use in isochrones. However, because of the lingering disagreements
between solar abundance tabulations, we will examine $Z \approx 0.017$
as a possible alternate composition for NGC 6811 stars.

\subsection{The Color-Magnitude Diagram}

To make the most effective use of the information in the CMD, we have
vetted cluster members using proper motions and radial velocities from
the literature where possible. We have attempted to use a uniform set
of photometry covering as much of the field as possible, using the
\citet{janes} dataset supplemented for a few stars with \citet{glush}
photometry as well as our own. (The comparisons in \S \ref{gbphot}
show that these datasets are have the same photometric zeropoints to
within 0.02 mag.) We will primarily focus on the turnoff region ($V
\la 13.8$) for which proper motion membership probabilities are
available from \citet{sanders6811} or \citet{khar13}\footnote{We
    only use the ``kinematic'' or proper motion membership
    probabilities from this study, and not the probabilities based on
    spatial position or photometry. It should also be noted that
    \citet{khar13} membership probabilities were calculated in a
    different way than \citet{sanders6811}, who accounted for the
    distribution of non-members in the field. Thus, the probabilities
    are not precisely comparable.}, or have been calculated by
  \citet{dias14} or can be calculated from the Fourth US Naval
  Observatory CCD Astrograph Catalog \citep[][UCAC4]{ucac4}. Because
  of occasional large disagreements between studies, we do not
  eliminate any stars from consideration if they have proper motion
  membership probabilities $P_\mu > 50$\% in any of the studies.

Radial velocities in the field have been published for stars in the
vicinity of the giant branch (10 stars, \citealt{merm}; 60 stars,
\citealt{glush}) and for stars near the giant branch and upper main
sequence (64 stars, \citealt{frinch}; 15 stars, \citealt{mz}). Radial
velocity membership measurements by S. Meibom for 5 red clump stars
were presented by \citet{stello}, and the APOGEE project
\citep{apogeespec,dr12} has also released spectroscopic observations
for 2 AGB and 4 red clump stars we identify below. In general, there
are only a handful of observations per star from these
sources. S. Meibom has an extensive database of unpublished radial
velocity measurements that we use here in a limited fashion. We report
the radial velocity membership as ``Y'' if there is no significant
variation in all of the measurements or if an orbital solution allows
us to identify that the system velocity is consistent with the cluster
mean ($\sim 7$ \kms). If there was significant radial velocity
variation, but the system could potentially be a cluster member, we
identify the star with a ``B'' (for binary candidate). If there is
little or no chance that the system is a member of the cluster, we
identify it with an ``N''. Without an even more extensive velocity
survey of the cluster that can reliably determine systematic
velocities of multiple star systems, the velocities will not give a
definitive determination of cluster membership, and we will be limited in
our ability to use them to identify where the single star locus should
be in the CMD.
For the stars with $V < 12$, both radial velocity and proper motion membership
information is usually available.

Sky position can also be used as circumstantial evidence of
  membership --- the cluster stellar density profile becomes so small
  beyond about two effective radii that the default should be that
  these are not cluster members. However, it is reasonable to expect
  that there are a small number of stars in the cluster outskirts, and
  we do indeed find stars where all other indicators are for
  membership. In Tables \ref{giants} -- \ref{misc}, we give a simple
  position membership indication (``Y'' or ``N'') based on whether
  each star lies within or without $13\arcmin$ from the cluster
  center.

Below is a short discussion of the stars in some
important groups.

\subsubsection{Red Clump Stars and Asteroseismology}

Masses derived from eclipsing binaries can benchmark masses inferred
from asteroseismology.  For star clusters in the \kep field, the
asteroseismic targets with detectable solar-like oscillations are on
the giant branch and in the red clump, while eclipsing binaries have
only been found on the subgiant branch and fainter. However, when used
in concert with theoretical models, the eclipsing binaries do make it
possible to predict giant star masses. In the asteroseismology arena,
there is still need to validate masses against accurate and precise
measurements. In the so-called direct method, which is likely to have
the widest application in astronomy in the near future, masses are
derived from scaling relations using the observables $\Delta \nu$ (the
frequency separation of overtone modes) and $\nu_{\rm max}$ (the
frequency of maximum power) along with $T_{\rm eff}$.  For the two
older open clusters in the \kep field, preliminary comparisons have
been made between asteroseismic masses and predictions from
theoretical models anchored with eclipsing binary measurements. In
both the $\sim8$ Gyr cluster NGC 6791 \citep{brogaard2} and the
$\sim$2.5 Gyr cluster NGC 6819 \citep{sand6819}, predicted red giant
masses are lower than calculated from asteroseismic scaling
relations. However, there is strong evidence that the offset between
binary-star and asteroseismic-inferred masses is reduced significantly
when stellar model-informed corrections are applied to $\Delta \nu$ in the
seismic scaling relations \citep{brogaard16,brogaard15,sharma}. An examination of NGC 6811's red clump stars
allows us to extend the comparison to more massive stars. The
clump stars in NGC 6811 did not undergo a degenerate flash at helium
fusion ignition --- this starts to happen in star clusters with ages
near 1.5 Gyr \citep{girardi,girardi09}. Thus we will be probing the
so-called secondary red clump (hereafter RC2) stars that have a
different evolutionary history.

We therefore re-examined stars in the cluster field with photometry
putting them in the vicinity of the red clump. Most of these have been
discussed by \citet{stello}, \citet{janes}, and \citet{mz}, and their
vital data is summarized in Table \ref{giants}. According to proper
motion measurements, four candidates are unanimously cluster
  members, while the remaining four have membership probabilities less
  than 50\% according to one or two studies (most often
  \citealt{khar13}), while the others indicate high likelihood of
  membership.
Available radial velocity information indicated membership for all of
the candidates with the exception of KIC 9655101 in \citet{frinch},
which may be a single-lined spectroscopic binary \citep{mz}.  KIC
9655167 is also identified as a single-lined spectroscopic binary by
\citet{mz} and \citet{merm}, and shows radial velocity variability in
the survey by S. Meibom. \citet{glush} identify KIC 9532903 as
velocity variable although \citet{mz} do not see variability in their
two measurements.

In addition to providing a way of estimating the stellar masses and
radii, asteroseismic measurements can be used to examine cluster
membership of individual stars, as shown by \citet{stello}. This
involves exploiting the correlation between the asteroseismic
observables and apparent magnitude for stars with very similar mass
and effective temperature. The asteroseismology of seven stars (one
early AGB and six red clump) have been discussed by different authors
\citep{stelloa,stello,hekker,corsaro}, but we have redone the
asteroseismic analysis of the nine NGC 6811 candidates using all
available long-cadence data from \kep (starting with either quarters 0
or 1 and reaching quarter 17) in Data Release 23. The longer time
baseline compared to previous studies has allowed us to reduce the
previously determined uncertainties on the asteroseismic parameters
$\Delta \nu$ and $\nu_{\rm max}$ by almost a factor of $1.5 - 2$ relative to
\citet{corsaro}. The sample of stars and the precision of the asteroseismic
observables will therefore be about the best currently possible for
this cluster.

PDC pipeline data in contiguous sections were fit with a low-order
polynomial in order to remove instrumental trends and
quarter-to-quarter zero-point differences. The time series were
analyzed using the pipeline method of \citet{huber09}, which provided
the values of $\Delta \nu$ and $\nu_{\rm max}$ including uncertainties
derived following \citet{huber11}. The observables are presented in
Table \ref{giants}.

\begin{figure}
\includegraphics[scale=0.3]{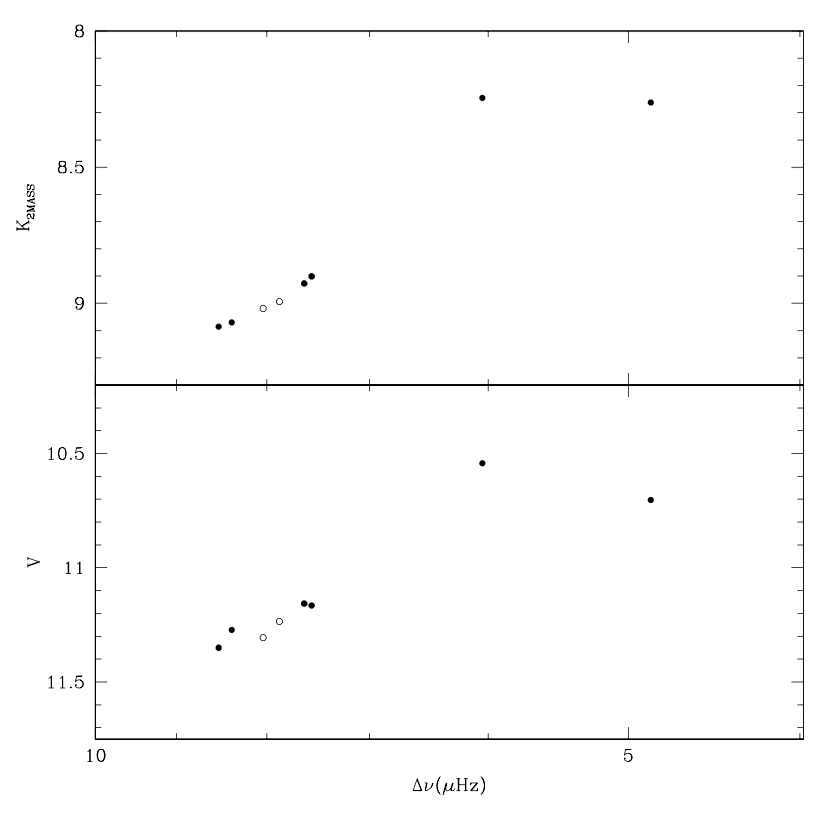}
\caption{Apparent magnitude versus asteroseismic large frequency
  spacing $\Delta \nu$ for giant star candidates in NGC
  6811. Single-lined spectroscopic binaries are shown with open
  symbols.\label{asteromem}}
\end{figure}

One candidate (KIC 9897838) had no detectable oscillations, and along
with its bluer color, we believe it is likely to be a field star. The
rest of the candidates show a tight correlation between $\Delta \nu$
and either $K$ or $V$ magnitude with the possible exception of KIC
9409513 (see Fig. \ref{asteromem}). We also compared predicted and
observed values of $\Delta \nu$ and $\nu_{\rm max}$ in the same manner
as \citet{stello}. Those authors employed scaling relations cast using
powers of mass, effective temperature, and luminosity to predict the
asteroseismic observables from stellar photometry assuming a common
mass for the giants, and a distance modulus and reddening for the
cluster.  This method indicates that all of the remaining red clump
stars are highly probable cluster members, while the two potential AGB
stars deviate more from the predicted values. The brighter
one (KIC 9409513) deviates the most, but its photometry could possibly
be explained by a blend with a bright main sequence star. The star is
classified as a radial velocity member, but some individual APOGEE
elemental abundances deviate mildly ($\sim0.10-0.15$ dex) from those
of other clump stars. The asteroseismic scaling relations imply that
KIC 9409513 is larger than the cluster clump stars ($8-9 \rsun$),
consistent with having evolved past its core helium burning
phase. This conclusion is insensitive to error in $T_{\rm eff}$
resulting from a stellar blend that affects the color.

Based on the weight of the evidence (including CMD position), KIC
9716522 is an unambigous AGB star cluster member, KIC 9409513 is a
possible additional cluster AGB star, and the remaining stars are
helium burning clump star members of the cluster. The only
dissents come in the proper motions for four stars from \citet{khar13},
as mentioned earlier.

Determination of asteroseismic masses depends on the accuracy and
precision of the photometric data used to determine the parameters
$\nu_{\rm max}$ and $\Delta \nu$, and of the method used to determine
$\teff$. Our calculations used the maximal \kep datasets to reach
precisions better than the previous determinations of \citet{stelloa}
and \citet{mz}. Because the uncertainty on any individual star's mass
is relatively large (no less than about 5\% even in our study),
cluster giants are generally used as an ensemble to improve the
precision of the mass measurement, which in turn provides the
asteroseismic constraint on the age of the cluster (e.g.,
\citealt{basu}). We have also improved the situation here by
identifying all of the cluster giants: both \citeauthor{stelloa} and
\citeauthor{mz} studied 5 giants. As for $\teff$, \citeauthor{stelloa}
used photometric color-$\teff$ relations, and so were subject to
systematic errors involving the poorly known cluster reddening and
metallicity at the time. \citet{stelloa} derived a mean mass of
$2.35\pm0.04 \msun$ from a shorter dataset for five stars in this
sample. \citet{mz} used spectroscopic $\teff$ and derived a mean
mass of $2.12\pm0.14 \msun$.

After calculating $\teff$ using the more recent low value for
reddening [$E(B-V)=0.07\pm0.02$; see \S \ref{chemred}] and $V-K_s$ colors
\citep[employing][]{rm05}, the mean mass determined from the asteroseismic
relations is $\bar{M} = 2.28\pm0.03\pm0.04 \msun$, where the
uncertainties are the error in the mean and the systematic uncertainty
due to the reddening, respectively.
The change in the reddening value largely accounts for the difference
with the \citet{stelloa} value.  As seen in Table \ref{giants}, our
photometric temperatures agree quite well with the spectroscopic
values of \citet{mz}. In addition, we found that the three stars (KIC
9655167, 9716090, and 9776739) with the lowest masses calculated by
\citet{corsaro} from the asteroseismic observables are here found much
closer to the mean for the other clump stars with these new
values. (This is due to large changes of 4-8\% in either the $\nu_{\rm
  max}$ or $\Delta \nu$ observables for these stars.)
The improvements underline the idea that the measurements for giants in the
open clusters NGC 6791 and NGC 6819 should be revisited with the full
\kep datasets. 

The reader should keep in mind that the clump phase, unlike the red
giant branch (RGB) phase, is long enough that a mass range of at least
a few hundredths of a solar mass is expected among the stars there,
according to models. When that is coupled with a small sample of red
clump stars as it is in NGC 6811, the average mass does not constrain
the cluster age as strongly as it would with a comparable number of
giants in an older cluster.  Asteroseismic radii have higher precision
than asteroseismic masses ($2-3$\% versus $6-7$\% random uncertainty),
and there is some dependence of the minimum clump star radius on
age. For the six clump stars, we find weighted average values $\bar{M}_{RC} =
2.24\pm0.03\pm0.04 \msun$ and $\bar{R}_{RC} = 8.59 \pm 0.12 \pm0.05 \rsun$,
where the quoted uncertainties are again random and systematic (due to
reddening).

\begin{figure}
\includegraphics[scale=0.3]{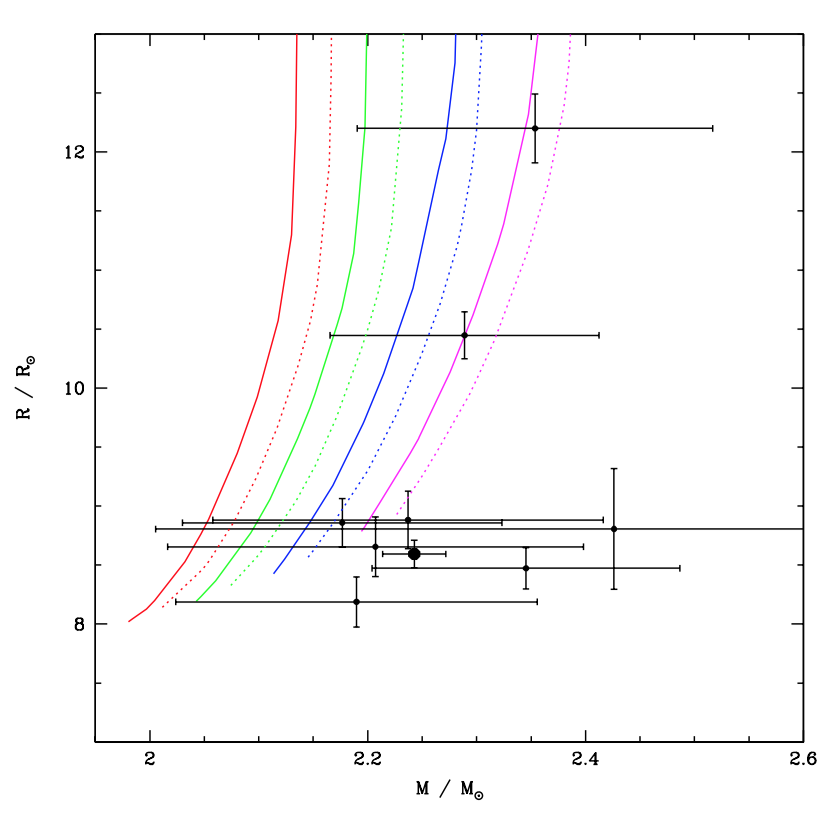}
\caption{Asteroseismic mass-radius plot for helium-burning stars in
  NGC 6811, with the weighted average for the clump stars ($R < 9 \rsun$) shown
  with the large point. Models have ages of 0.9, 1.0, 1.1, and 1.2 Gyr
  (from right to left) for PARSEC \citep{parsec} isochrones, and have
  been restricted to helium burning phases for clarity. Solid lines
  use $Z \approx 0.0137$, and dotted lines use $Z \approx 0.0167$.
\label{mrclump}}
\end{figure}

Although there are limited choices for comparisons with models, it is
important to compare the asteroseismic inferences with predictions
from stellar evolution utilizing our knowledge of the characteristics
of the eclipsing binary stars. In Fig. \ref{mrclump}, there is
reasonable agreement with an age near 0.9 Gyr using PARSEC models.
This age is younger than implied by the eclipsing binary, which can be
interpreted as saying that the asteroseismic masses for the clump
stars are larger than expected from the main sequence.
There are potential systematic effects in the models that can affect
this comparison, however.  The reddening affects the asteroseismic
masses more than the radii, so that a lower reddening moves the stars
toward larger ages.  Metallicity errors also affect the model
masses more than radii (compare the solid and dotted
lines in Fig. \ref{mrclump}), in the sense that higher metallicity
leads to larger ages. On the physics side, a larger mixing length
parameter, for example, can reduce the computed radii of red clump
stars \citep{yang}. The scaling relations used to calculate the radii
may also need to be modified to apply accurately to red clump
stars. Because the $\Delta \nu$ parameter relates to sound speed
within a star, and because red clump stars have different structure
than the red giant branch stars to which the scaling relations are
usually applied, models indicate that $\Delta \nu$ is systematically
higher for clump stars compared to giants with the same mean density
\citep{miglio}. However, models indicate that a clump star's $\Delta
\nu$ value is a more {\it accurate} indicator of its average density
than it is for a giant \citep[for example, see Fig. 4 of][]{sharma},
and corrections to the scaling relations are more important for RGB
stars.  For RGB stars that have been studied in other \kep clusters,
there are indeed signs that uncorrected asteroseismic masses are
overestimated \citep{brogaard2,sand6819}. With a greater degree of
constraint on the age of NGC 6811 from further study of the binary
stars, the need for such corrections will be more strongly addressed.

\subsubsection{The Main Sequence Turnoff and Pulsating Stars}\label{to}

Because a detailed membership study has not been completed for stars
around the cluster turnoff, we need to identify the ones that should
be used to delineate the evolutionary sequence for single stars. As a
first step, we briefly review evidence for binarity among the brighter
cluster stars in order to identify systems that could confuse the
location of the single-star evolutionary sequence.  A handful of
binaries can be identified from their eclipses \citep{kebs2}. In the
cases below, the proper motion indicators agree on cluster membership,
but radial velocities were discrepant from the mean. They are 
therefore possible cluster members.  KIC 9655187/Sanders
110 and KIC 9655346/Sanders 127 are known short-period eclipsing
binaries, easily explaining their radial velocity values.
KIC 9716456/Sanders 159 has no radial velocity measurements, but is
also a short-period eclipsing binary.  KIC 9533489/Sanders 256 is a
long period (197.15 d) eclipsing system with only one obvious eclipse
that falls near the bluest part of the cluster main sequence, so that
there is a good chance of cluster membership if the primary star
dominates the photometry. KIC 9656397 was not identified in the Kepler
Eclipsing Binary Catalog, but it shows two distinguishable eclipses
during each 204.74 d cycle, and we have unpublished radial velocities
showing that the system velocity matches the cluster mean. Finally,
there are two fainter binaries without proper motion measurements. KIC
9655129 is a short-period eclipsing system, and KIC 9837544 is a
longer period (71.662 d) binary. In both cases, the photometry puts
them well to the red of the cluster main sequence, in spite of some
support for membership from proper motions.

\begin{figure*}
\includegraphics[scale=0.4,angle=-90]{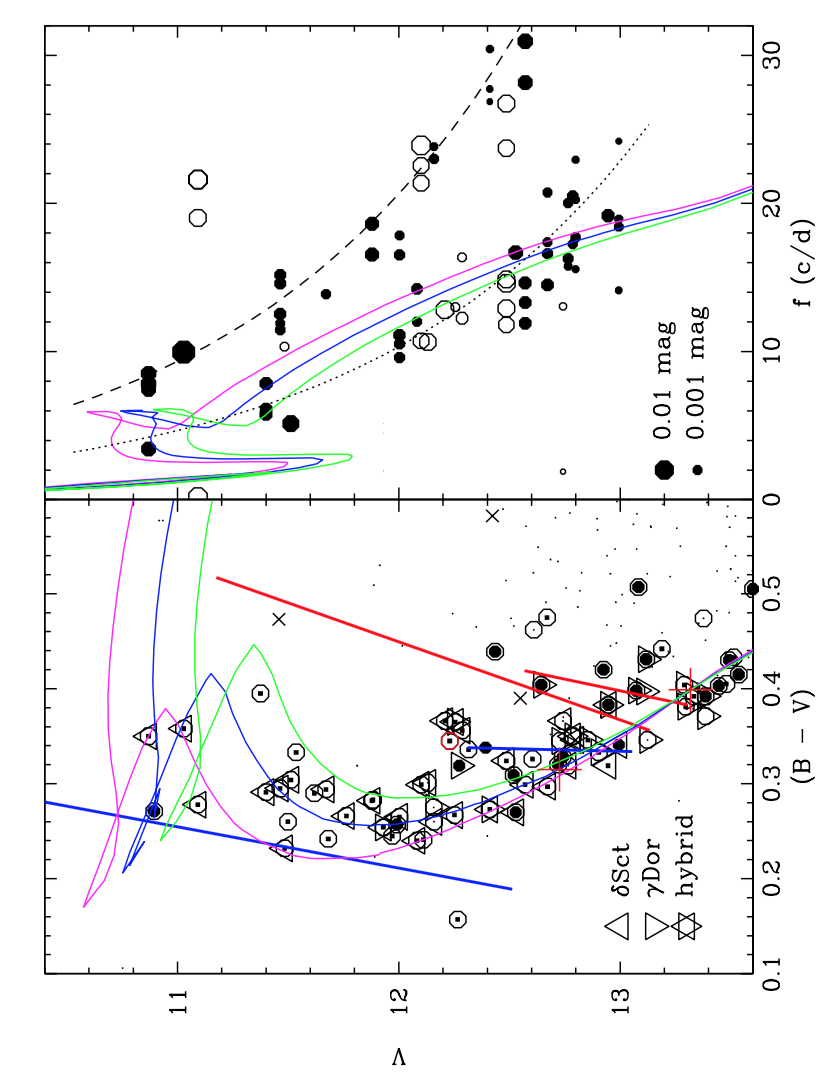}
\caption{{\it Left panel:} Color-magnitude diagram for pulsating stars
  of $\delta$ Sct, $\gamma$ Dor, or hybrid types in NGC
  6811. Empirical instability strip boundaries are from \citet{uytter}
  translated using $(m-M)_V = 10.33$ and $E(B-V)=0.07$. The components
  of KIC 9777062 are shown with $+$, and radial velocity non-members
  are shown with $\times$.  Isochrones are from the PARSEC
  (\citealt{parsec}; $Z=0.0137$) sets for ages of 0.9, 1.0, and 1.1
  Gyr. {\it Right panel:} The highest amplitude modes ($>0.5$ times
  the amplitude of the strongest)
  $\delta$ Sct stars versus $V$ magnitude.  The dotted line is the
  period-luminosity relation of \citet{mcnamara}, and the dashed line
  is the same relation with the frequency doubled. The colored lines
  are the PARSEC isochrones with average density used to predict the
  fundamental mode frequency, assuming $Q = 0.033$ d. Likely
  photometric binaries are shown with open symbols.
  \label{cmdpuls}}
\end{figure*}

In NGC 6811, we can make use of an additional piece of circumstantial
evidence that can argue for cluster membership: pulsation. We roughly
translated the empirical boundaries of the $\delta$ Sct and $\gamma$
Dor instability strips shown in \citet{uytter} into the color
magnitude diagram using \citet{vandc} color-temperature relations,
assuming [Fe/H]=0, $E(B-V)=0.07$, and $(m-M)_V=10.33$. Although there
are significant uncertainties in all of the quantities involved in
making this translation (including the position of these empirical
boundaries in $\teff$ and log $g$), Figure \ref{cmdpuls} shows that
cluster stars at the turnoff and redder can realistically be expected
to be pulsating stars. This observation has been made previously by
\citet{frand} and \citet{luo}, but the existence of \kep photometry
allows us to identify pulsators at quite low amplitude levels.

$\delta$ Sct and $\gamma$ Dor stars in the NGC 6811 field have been
previously identified in ground-based \citep{vanc,luo} and \kep
\citep{deboss,uytter,tka} data. \citeauthor{uytter} focused on these
two groups throughout the \kep field, as well as hybrids showing
frequencies belonging to both types. \citeauthor{deboss}
probabilistically categorize variable stars based on Fourier
decomposition of \kep quarter 1 data, though those authors found in a
previous paper \citep{deboss09} that their $\delta$ Sct and $\beta$
Cep classifications overlapped frequently --- most of the stars they
identify as $\beta$ Cep in the NGC 6811 field are actually $\delta$
Sct. To produce a clean list of pulsators, we therefore examined the
\kep light curves of all of the stars near the turnoff and identified
the type of variation using the frequency spectra.

To compute spectra, we downloaded the largest contiguous set of short
cadence data when available, and long cadence data when not. Either
type of data allows us to examine $\gamma$ Dor frequencies ($f < 5$
c/d or 58 $\mu$Hz). Long cadence data probes $\delta$ Sct below the
Nyquist frequency ($f \approx 24.5$ c/d or 0.14 mHz), with some strong
super-Nyquist frequencies potentially reflected in the spectrum as
well \citep{murph13}. We used Period04 software \citep{p04} to
prewhiten the data successively after identifying and subtracting the
highest-amplitude frequency remaining. Table \ref{pulsate} lists the
stars with observed pulsation. A comparison of Tables \ref{pulsate}
and \ref{misc} shows that stars with detectable pulsation constitute
the majority falling within the expected instability strips.

We plan to discuss the spectra in more detail in an upcoming
paper, but we comment on a few aspects of the pulsations that are
relevant to establishing their cluster membership.  Our understanding
of the frequency structure of $\delta$ Sct stars is still limited, but
gross features may reveal important information. For example, we
expect the radial fundamental mode (if excited) to be near the lower frequency
limit for $\delta$ Sct stars, although this is complicated by rapid
rotation and the possible excitation of gravity or mixed modes.
Based on dynamical timescale arguments, we should also expect that
less luminous stars in NGC 6811 will be significantly more dense and
would tend to have their pulsation modes shifted to higher frequencies ($f
= \sqrt{\rho / \rho_\odot} / Q$, where $Q$ is the pulsation constant
for the mode).  We plot the strongest observed frequencies for each
$\delta$ Sct star
as a function of $V$ magnitude in the 
right panel of
Fig. \ref{cmdpuls}. A large proportion of the strongest frequencies
fall approximately where predicted
for the radial fundamental mode ($Q = 0.033$ d) using average
densities derived from fitted isochrones, or using the period-luminosity
relation for the fundamental mode of high-amplitude $\delta$ Sct stars
with Hipparcos parallaxes \citep{mcnamara}. These stars are found from
the faint red end of the sample up to near the estimated blue end of
the instability strip, consistent with the expectation that the
fundamental mode is not excited at the blue edge \citep{pamy}. We
believe this, along with the proper motions and color-magnitude positions of the stars,
is good evidence that these stars are cluster members.

Many of the brighter stars have pulsation frequencies that are
inconsistent with the radial fundamental mode or the first overtone
radial mode (theoretically expected to have a period ratio of $P_1 /
P_0 = f_0 / f_1 \approx 0.772$ with the fundamental). These
frequencies can be found near the dashed line in the 
right panel of
Figure \ref{cmdpuls}. Although most of the faintest stars only have
long cadence \kep observations, and high frequency pulsations would be
above the Nyquist limit, several stars with $V > 12.6$ have identifiable
frequencies there. 

Based on these observations, we derived an apparent distance modulus
by fitting these data with the \citet{mcnamara} period-luminosity
relation. We restricted the sample to stars with strong frequencies
nearest the expected value of the radial fundamental mode, although it
is likely that physical processes (like rotational splitting for the
higher frequency modes) would shift the frequencies that are actually
excited. We note that binary companions to these stars would tend to
bias their $V$ magnitudes higher than they should be, and could cause
us to underestimate the distance modulus. We compare the resulting
distance modulus $(m-M)_V = 10.37 \pm 0.03$
with the results of the binary modeling in the next section.

In the remainder of this section, we discuss the membership of stars in
different portions of the main sequence in preparation for using the
color-magnitude diagram in model comparisons and age determination.

Stars brighter than the most heavily populated part of the main
sequence ($V \lesssim 11.4$) are potentially important because they
could trace the evolution at the time of core hydrogen exhaustion
and constrain physical processes (like convective core overshooting)
that affect the evolution. At the same time, unidentified binaries or
blue stragglers can easily complicate model comparisons. Among these
bright main sequence stars, the most important is probably
KIC 9777532/Sanders 247 because it has unambiguous evidence of
membership and no evidence of binarity. If its photometry is truly
representative of a single star, it probably resides on the blue hook
at the end of the main sequence isochrones. The other stars that have
radial velocity membership information in addition to proper motions
are KIC 9655177/Sanders 108, KIC 9716385/Sanders 136, KIC
9655543/Sanders 172, and KIC 9594857/Sanders 205. Three of these stars
(S108, S136, S205) have been identified as $\delta$ Sct variables,
and additional information on their interiors might be gleaned from
their pulsations in the future. But among these stars, S136's
pulsations stand out with an unusually high frequency in the $\delta$
Sct regime ($f \approx 22$ cycles per day). If its photometry is the
result of being part of a nearly equal-mass binary, its pulsation
frequency would be more in line with fainter cluster stars.
Four radial velocity measurements to date (\citealt{frinch}; Meibom)
show a large amount of velocity variability consistent with this.
More velocity observations are warranted for this star. S205
shows signs of velocity variability in the observations of \citet[2
  observations]{mz} and \citet[1 observation]{frinch}, indicating that
it is a likely single-lined spectroscopic binary (SB1). S108 only has
one radial velocity observation \citep{frinch}, but it deviates from
the cluster mean by about 15 km s$^{-1}$.  So S247 appears to be the most
likely single cluster member among these stars.

For stars on the main sequence above the cluster turnoff ($11.4 < V <
12$), there is a fair amount of scatter in the CMD that obscures
the single star evolutionary sequence. In this group we can
only identify one star (KIC 9716220/Sanders 113) that is a member by
all criteria (radial velocity, proper motion, and pulsations), although
it shows signs of velocity variability. Three other stars (KIC
9655055/Sanders 86, KIC 9655422/Sanders 144, and KIC 9655444/Sanders
149) are proper motion members and pulsators, but each shows velocity
variability. Two additional stars (KIC 9655005/Sanders 77 and KIC
9655382/ Sanders 134) are proper motion members and lack detectable
pulsations. Both appear to be radial velocity members, but KIC 9655005
shows velocity variability while KIC 9655382 does not.
The remainder of the stars in this range are
eclipsing binaries, have radial velocities that deviate from the cluster mean
and could be spectroscopic binaries (S75, S87, S121, S161, S188, S230), or
have detectable light travel time effect on $\delta$ Sct pulsations (S87;
detected using the method of \citealt{murphy}). Thus, we will restrict
ourselves to at most 6 stars in this range of magnitudes.  More precise CMD
age constraints will require radial velocity study of candidate members here
because this is where the isochrones start to separate most strongly.

Below the turnoff to the faint end of the instability strips for
$\delta$ Sct and $\gamma$ Dor stars ($12 < V \lesssim 13.4$), there
are fewer published radial velocities for making membership and
binarity judgements, but we can make better use of CMD position to
eliminate binaries with bright secondary stars.
The faintest $\delta$ Sct stars that are likely members have $V < 13$,
while we find $\gamma$ Dor stars down to $V \approx 13.4$.

\begin{figure}
\includegraphics[scale=0.3]{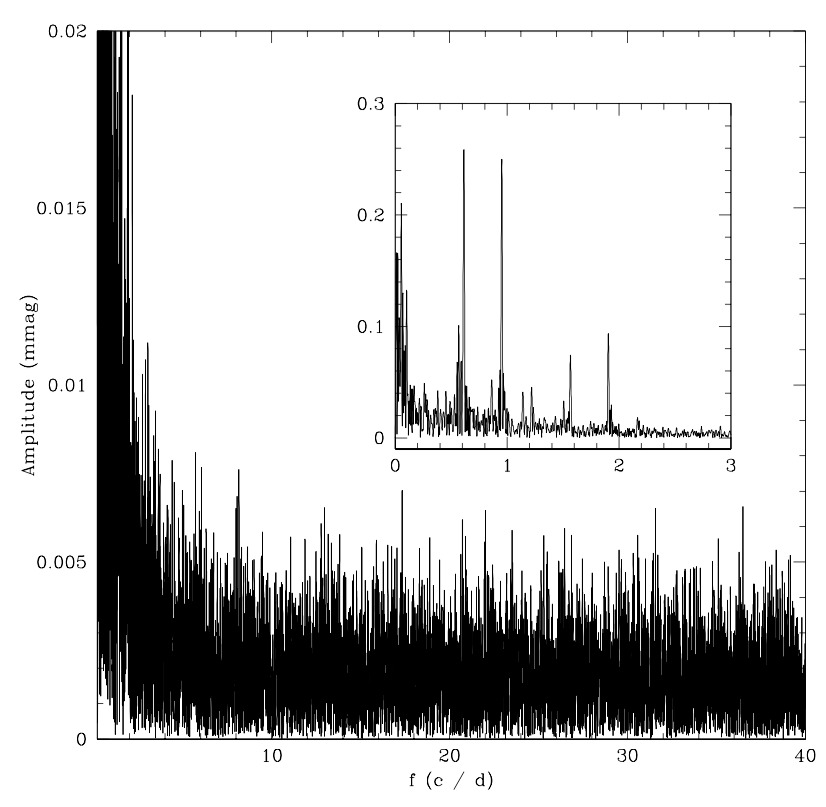}
\caption{The amplitude spectrum for KIC 9777062 after subtraction of
  model eclipses from the light curve. The inset shows a zoom on the
  detected low frequencies.\label{s195spec}}
\end{figure}

Before moving on to a discussion of age and model comparisons, we
return to the stars in the eclipsing binary. The deconvolved
photometry of the two components of KIC 9777062 are plotted in
Fig. \ref{cmdpuls}, and their positions are consistent with other
likely cluster single-star members. Because the stars appear to fall
within the $\delta$ Sct and $\gamma$ Dor instability strips
respectively, we searched for evidence of pulsation in the light
curve. We used short cadence \kep data from the entirety of quarter
10, and subtracted the best fit model of the eclipses. The spectrum
(Fig. \ref{s195spec}) shows no evidence of $\delta$ Sct frequencies
($> 5$ cycles per day), but there are some low frequencies that may be
associated with $\gamma$ Dor pulsation. The amplitudes are about 0.25
millimag for the two strongest modes (0.6133 and 0.9525 c/d). The
small amplitude of the modes (about an order of magnitude smaller in
size than the secondary eclipse) means that they affect the eclipse
light curves minimally. The pulsations probably result from the
secondary star because the primary star appears to be well outside the
estimated position of the $\gamma$ Dor instability strip. As the
pulsations are coherent, we investigated whether frequency modulation
due to the binary orbit could be measured and used to identify which
of the binary components is oscillating. Because we have characterized
the binary orbit, we can predict the amplitudes of the sidelobes of
the dominating observed oscillation modes in the power spectrum
\citep{fm} and these are far too low to be measurable.  We also tested
whether any phase shift could be measured \citep{murphy} when fitting
frequencies of the strongest modes to the light curve and isolating
different orbital phases in the light curve, but we did not find any
signal. However, we note that during the primary eclipse the light
curve with the best-fit model subtracted show slightly increased
scatter compared to the out-of-eclipse variation, which we take as
indication that the pulsations originate from the secondary component.
Although the primary star does not appear to be a pulsating star
itself, the very precisely determined masses and radii for both stars
can be used to predict masses and radii for other NGC 6811 pulsators.

\subsection{The Cluster Age and Distance Modulus}\label{overshoot}

To make the connection between cluster age and stellar
characteristics, theoretical models are necessary, and
systematic differences in the input physics or chemical composition
will affect the inferred age. Table \ref{isotab} summarizes some of
the more important physics inputs in presently available model
sets. To highlight physics differences, we use a common set of ages
for all of the comparisons between sets of theoretical isochrones, and
very nearly the same heavy element abundance $Z$. It is important to
remember that there are potential systematic errors in our choice of
$Z$, both due to uncertainties in the measured abundances relative to
the Sun, and due to continuing uncertainty over the Sun's heavy
element abundances. Even when forcing the models to have a common
heavy element abundance, different helium abundances will be present
in the models, both because of the need to calibrate the models to the
Sun for the included physics and because of different assumptions
about the galactic helium enrichment rate (usually parametrized as
$\Delta Y / \Delta Z$). For the models used here, the differences
between the highest and lowest abundances are modest (0.02 in $Y$).

\begin{deluxetable*}{lrrccrc}
\tablewidth{0pt}
\tablecaption{Summary of Relevant Inputs for Model Isochrone Sets}
\tablehead{\colhead{Isochrone} & \colhead{$Z$} & \colhead{$Y$} &
  \multicolumn{2}{c}{Diffusion?} & \colhead{Overshoot} & Mixing Length\\
  & & & \colhead{$Y$} & \colhead{$Z$} &
  \colhead{$\lambda_{OV}/H_P$} & \colhead{$\alpha_{ML}$}}
\startdata
Dartmouth & 0.01375 & 0.2659 & Y & Y & 0.20 & 1.94 \\
PARSEC & 0.01370 & 0.2726 & N/Y\tablenotemark{a} & N/Y\tablenotemark{a} & $\sim0.25$ & 1.74 \\
Victoria-Regina & 0.01250 & 0.2629 & N & N & \tablenotemark{b} & 1.90 \\
Yale-Yonsei & 0.01370 & 0.2574 & Y & N & 0.20 & 1.74
\enddata
\label{isotab}
\tablenotetext{b}{PARSEC isochrones assume no diffusion for a star
  with the mass of the primary star in KIC 9777062, but 
  diffusion acting for a star with the mass of the secondary.}
\tablenotetext{b}{Victoria-Regina models use a different algorithm for
  convective core overshooting, so it is difficult to directly compare
  to the other sets.}
\end{deluxetable*}

Because the stars in the eclipsing binary KIC 9777062 are relatively slow
rotators compared to the majority of single stars in their mass range,
diffusion may play an important role in our discussion.
The Victoria-Regina isochrones employed here \citep{vr} do not include
diffusion, although their more recent models have implemented helium
diffusion. For the PARSEC isochrones \citep{parsec}, diffusion is
turned off if the star is more massive than $1.60 \msun$ (which leaves
out the primary star and turnoff stars in NGC 6811) or when the mass
of the surface convection zone falls below $0.005 M$, where $M$ is the
mass of the star. For the Dartmouth isochrones \citep{dsep}, diffusion
is inhibited in the outermost $0.005 \msun$, and ramped up to full
strength $0.01 \msun$ below the photosphere \citep{chaby}. For the
Yonsei-Yale isochrones \citep{yy}, helium diffusion is allowed to run
at full strength.

As for uncertainties in convection physics, we briefly mention core
overshooting and the mixing length. Core overshooting generally does
not have a significant effect on observable properties until stars are
nearing hydrogen exhaustion: a larger amount of overshooting prolongs
a main sequence star's life and affects how rapidly the star needs to
adjust to hydrogen exhaustion. Differences here affect comparisons in
the CMD at the brightest part of the main sequence and on the subgiant
branch, but should not affect the $M-R$ comparison for our binary. The
mixing length ($\Lambda = \alpha_{ML} H_P$, where $H_P$ is the
pressure scale height) has its largest effects on the properties of
surface convection zones. Although the stars near the turnoff of NGC
6811 (including the stars of our binary) should have very low surface
convection zone masses, the evidence is that surface convection is
still present in the fainter half of the instability strip at least
and that the convection might be more turbulent than for cooler stars.

\subsubsection{Mass-Radius ($M-R$) Comparison}

\begin{figure}
\includegraphics[scale=0.3]{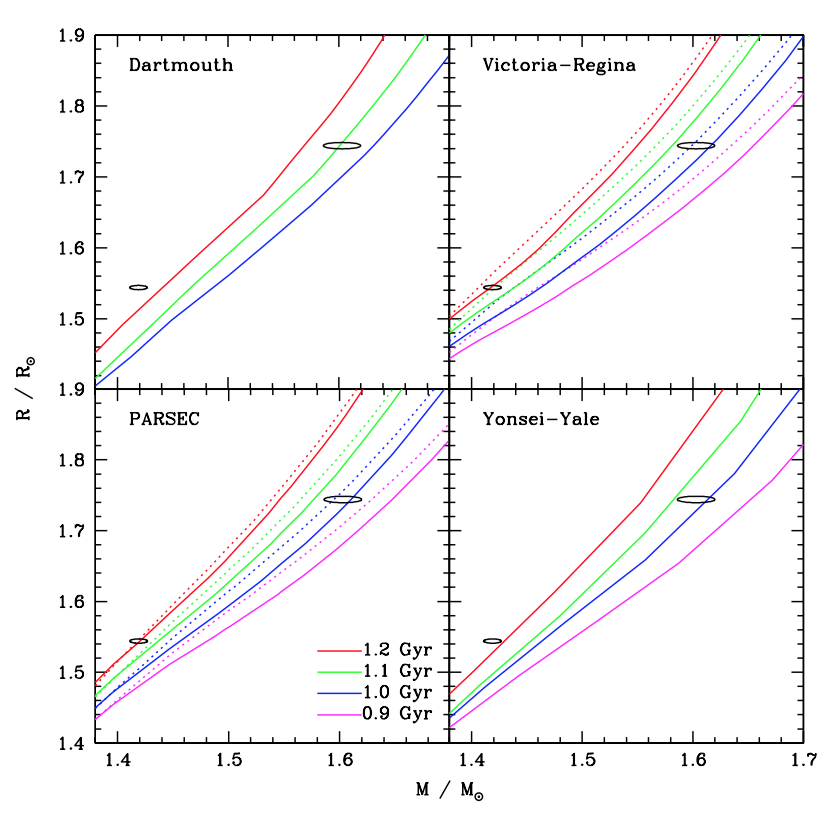}
\caption{Mass-radius plot for the members of KIC 9777062 with $1\sigma$ uncertainty ellipses. Models have
  ages of 0.9, 1.0, 1.1, and 1.2 Gyr (from bottom to top) for
  Victoria-Regina \citep{vr}, PARSEC \citep{parsec}, and Yonsei-Yale
  \citep{yy} models, while the Dartmouth \citep{dsep} models have ages
  of 1.0, 1.1, and 1.2 Gyr. Solid line isochrones use $Z \approx
  0.0137$ in all panels but Victoria-Regina. In the PARSEC panel, the
  dotted lines show $Z = 0.0167$. In the Victoria-Regina panel, our
  preferred $Z$ is midway between the solid and dashed lines ($Z =
  0.0125$ and 0.015, respectively).
\label{tomr}}
\end{figure}

As stated earlier, a comparison with models in the mass-radius plane
avoids common uncertainties that plague CMD comparisons: reddening,
distance, and color-$T_{\rm eff}$ transformation uncertainties.
Fig. \ref{tomr} shows the comparison with the components of KIC
9777062 in the $M-R$ plane. Isochrones from different groups look
similar, but there are interesting differences for each of the binary
components.  For the primary star, the Dartmouth isochrones
\citep{dsep} are noticeably offset toward smaller stars for a given
age compared to PARSEC \citep{parsec}, Victoria-Regina \citep{vr}, and
Yonsei-Yale \citep{yy} isochrones. For the secondary star, the
Dartmouth models again have the smallest radii at a given age, but the
Victoria-Regina models predict significantly larger radii than the
PARSEC and Yonsei-Yale models. The differences imply that there are
some significant differences in the input physics being used. We have
not found clear reasons to explain the differences between model
radii, however.  For example, if helium abundance was the dominant
influence on the model radii, the PARSEC models should have been
smallest and the Yale-Yonsei should have been largest, which is not
seen. The Dartmouth and Yonsei-Yale isochrones are the only ones that
incorporate helium diffusion at present, but these are not models with
the largest radii (due to higher opacity in the surface layers).
Because of the uncertainties in model radii, we will consider the
dispersion to be at least partially representative of the systematic
errors involved in deriving the age. This comes to approximately
$\pm0.04$ Gyr for the primary star, and $\pm0.10$ Gyr for the
secondary. If metallicity uncertainties are considered (as in the
PARSEC panel of Figure \ref{tomr}), there is an additional systematic
uncertainty of about 0.05 Gyr for both stars.

Although the mass and radius measurement uncertainties are larger for
the primary star in the binary, the two stars produce age estimates
with similar precision (about 0.05 Gyr) for a given set of isochrones
because the primary star's characteristics place it where there is
greater separation between isochrones. The age estimates produced by
the separate components of the binary, however, are different by about
$2-3\sigma$, with the secondary returning an age that is about
0.15-0.20 Gyr higher than that for the primary.

Regarding the different age estimates returned by the two binary components,
initial helium abundance can modify the shape of the isochrone
\citep{brogaard}. We do not consider this to be the most likely explanation
because NGC 6811 stars show elemental abundances that are otherwise quite
close to solar abundances (with the exception of barium; \citealt{mz}). 
If heavy element and helium abundances are connected ($\Delta Y /
\Delta Z$ is a constant, for example, as is often assumed), the small
super-solar metallicity would produce a minor enhancement in the
helium abundance ($\Delta Y \sim 0.001$) on its own. 

Main sequence stars having massive surface convection zones in
short-period eclipsing binaries have been shown to have radii that are
inflated by as much as 10\% above theoretical expectations
\citep{torres06}, possibly because magnetic activity helps inhibit
convective motions and makes the convection less efficient. It can be
argued that this kind of effect should be minimal in the KIC 9777062 system:
the orbital period is substantially larger than in affected binaries
($P \la 3.5$ d), and there should be a very low-mass surface
convection zone for the secondary star ($< 10^{-3} \msun$) and an
even less massive surface convection zone for the primary
star. Convection should be less efficient in the very shallow
convection zone of the primary due to radiative losses, and this is also
unlikely to help produce a smaller radius.

We can also examine whether there are effects on the radius related to
physics involved in the Am/Fm phenomenon.  One empirical approach that
can be used to examine this question is to look at precisely measured
eclipsing binary star systems that include Am/Fm stars. In the
interest of the paper's readability, we have put this discussion into
an appendix. Our conclusion is that
the properties of the secondary star in the $M-R$ plot for KIC 9777062 are likely to be the most
trustworthy for determining the age because it does not show the
extreme abundance anomalies that the primary star does. However, the
properties of the primary stars in this and other binaries are not
understood at present, and may be subject to physics that is not being
included in models. The model isochrones themselves do not agree at
the level we need to be able to derive a confident absolute age for
the cluster, and the differences between models cannot be attributed
confidently to one physics input (like diffusion) or another. We are
forced to conclude that more theoretical and observational effort needs
to be exerted to understand stellar radii in this range of masses.

Before moving on to color-magnitude diagram comparisons, we can use
the stellar radii in concert with a $\teff$ estimate to compute the
luminosity of each star. With an additional bolometric correction, the
absolute magnitude and apparent distance modulus can be calculated.
We utilized bolometric corrections in $V$ from \citet{cv14} along
with the model radii and spectroscopic temperature estimates from \S
\ref{constrain} to find $(m-M)_V = 10.47\pm0.09$ and $10.47\pm0.06$,
respectively for the primary and secondary
components. The largest contributor to the uncertainty in these
distance moduli by far is the temperature.
The averaged distance modulus is $(m-M)_V=10.47\pm0.05$.

The binary star distance measurements are modestly outside the
$1\sigma$ uncertainty ranges from previous determinations. \citet{mz}
derive $(m-M)_V = 10.29 \pm 0.14$ using a combination of spectroscopic
and asteroseismic information for 5 red clump stars, and our own
re-examination of the giant sample (omitting KIC 9409513) gives
$(m-M)_V = 10.31\pm0.04$, where the uncertainty estimate is the
standard deviation of the measurements (to better represent the likely
systematic nature of the errors in $\teff$ measurements).  Our analysis
of cluster $\delta$ Sct stars produced a measurement of
$10.37\pm0.03$. \citet{janes} quote a best estimate $(m-M)_V =
10.22\pm0.18$ from Bayesian comparison of isochrones with
color-magnitude diagrams, but there are systematic uncertainties
relating to the choice of an isochrone set. For example, they derive
$(m-M)_V = 10.13 \pm 0.19$ using Yale-Yonsei isochrones, but
$10.31\pm0.11$ using Padova isochrones. In addition, their analysis
found lower metallicities than have been found spectroscopically, and
this would lead them to overestimate the luminosity and derive
systematically smaller distance moduli.  At the same time, if our
spectroscopic temperatures for the binary are overestimated by less
than the uncertainties given above, our distance moduli would come
into better agreement.

\subsubsection{Color-Magnitude Diagram (CMD) Comparisons}\label{cmd}

\begin{figure}
\includegraphics[scale=0.3]{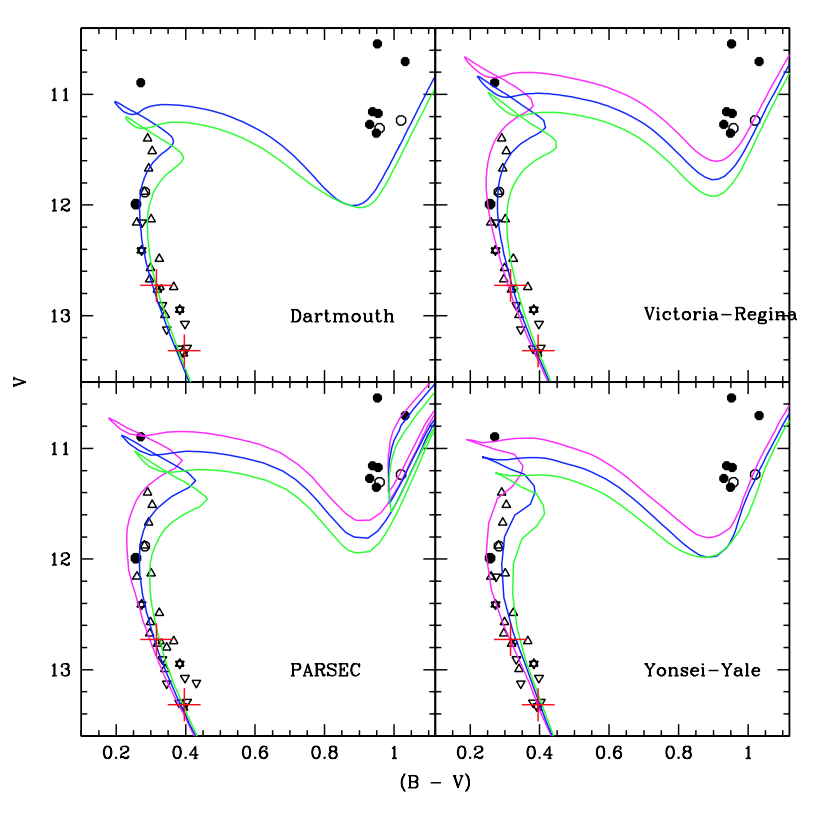}
\caption{Color-magnitude diagram comparison for fixed reddening
  [$E(B-V)=0.07$], distance modulus [$(m-M)_V=10.47$], and
  metallicity. Model isochrones have ages of 0.9 Gyr (magenta), 1.0
  Gyr (blue), and 1.1 Gyr (green) for Dartmouth ($Z=0.0147$),
  Victoria-Regina ($Z=0.0150$), PARSEC ($Z=0.0147$), and Yonsei-Yale
  ($Z=0.0147$) isochrones. Points have the same meaning as in
  Fig. \ref{cmdpuls}.
\label{redcmd}}
\end{figure}

Any comparison in the CMD using just the distance modulus and
reddening to shift the theoretical isochrones will also be subject to
issues involving inaccuracies in the color-$T_{\rm eff}$
transformations, but we do a preliminary comparison in
Fig. \ref{redcmd} using the distance modulus derived from the binary
system and a reddening that moves the main sequence into approximate
agreement with the fainter star in the binary. The four sets of
isochrones appear to agree on an age of $0.9-1.0$ Gyr, and this
includes the potentially single star at $V \approx 10.9$.
That star (KIC 9777532/Sanders 247) may be a legitimate cluster subgiant. If
so, it can give us good leverage on the age from the CMD. More
study is needed in order to determine whether its CMD position is
affected by an undetected companion, however.

Because we have mass measurements for the KIC 9777062 stars, we can
force isochrones to fit their CMD positions, thereby constraining
isochrone positioning further. This should also minimize the effects of
uncertainties in reddening, distance modulus, and color-temperature
relations. 
A first comparison to make is whether the models match the
observed magnitude separation for the measured star masses. The
observed $\Delta V = 0.586\pm0.021$ 
is smaller than the model predictions by about 0.07 mag,
and this cannot be explained by calculated uncertainties in the star masses or
reasonable uncertainties in the age or metallicity of the cluster. 
As discussed in the previous subsection, this may be
an indication that the physics implemented in the
model isochrones is not adequately describing what is occuring
in the primary star. Diffusion is a leading suspect in this
case because of the Am nature of the primary. We therefore
think that the most reliable comparison will be between the
model isochrones and the secondary star. If most of the
other stars near the turnoff of NGC 6811 do not show Am/Fm
characteristics, then our CMD comparisons should also be
valid. Tests of these assumptions are beyond the scope of
this paper, but the analysis of other eclipsing binaries in
the cluster would help. Lower-mass stars are expected to be
less affected by diffusion due to their larger surface
convection zones, and we predict that the properties of
these stars relative to the secondary of KIC 9777062 will
be in line with models.

\begin{figure}
\includegraphics[scale=0.3]{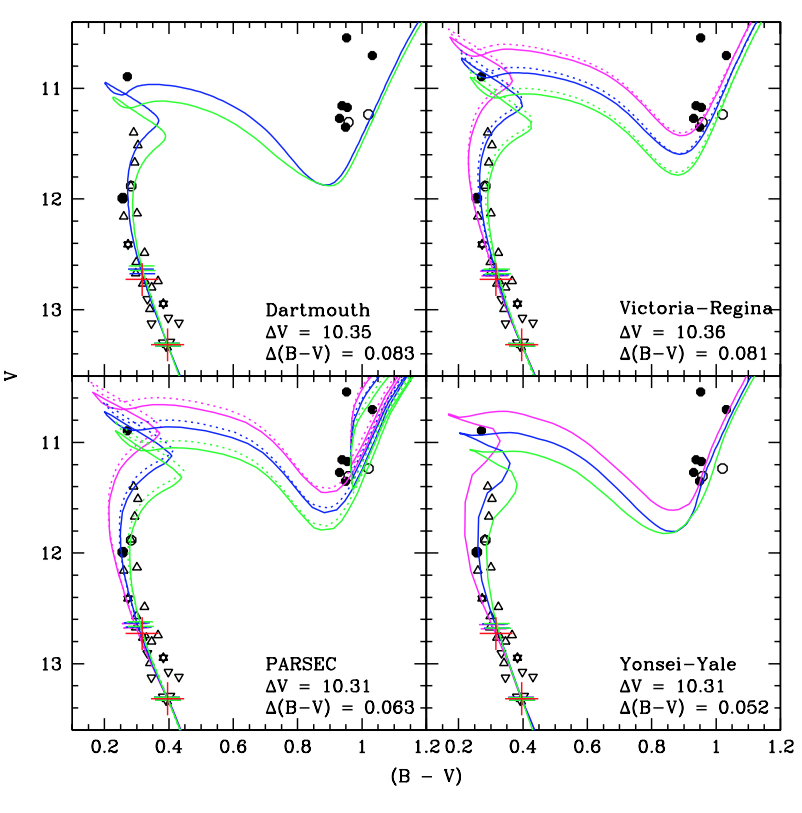}
\caption{Color-magnitude diagram for selected probable members of NGC
  6811 with isochrones shifted to match the mass and photometry of the
  secondary star of KIC 9777062. (Magnitude and color shifts are given
  in each panel.) All stars have proper-motion membership
  probabilities $P_{PM} > 0.50$ \citep{sanders6811}. Radial velocity
  members are shown with $\bullet$. Pulsating stars and the components
  of KIC 9777062 have the same symbols as in Fig. \ref{cmdpuls}.
Isochrones are from the Dartmouth (\citealt{dsep}; $Z = 0.0137$), 
Victoria-Regina (\citealt{vr}; $Z=0.0125$ and 0.0151 with solid and dashed lines), PARSEC (\citealt{parsec};
$Z=0.0137$, dotted lines show $Z=0.0167$), and Yonsei-Yale
(\citealt{yy}; $Z=0.0137$) sets for ages of 0.9, 1.0, and 1.1 Gyr
(magenta, blue, and green respectively).
The predicted position of the primary star relative to the
secondary for different isochrones are shown at $V \approx 12.6 - 12.7$.
\label{cmdfix}}
\end{figure}

Fig. \ref{cmdfix} shows a comparison of isochrones to cluster star
photometry for our restricted set of members that are least likely to
be binaries. For all of the isochrone sets, the bluest stars on the
upper main sequence imply an age just under 1.0 Gyr. In the case of
the Victoria-Regina, PARSEC, and Yonsei-Yale isochrones, the potential
subgiant at $V \sim 10.9$ also indicates an age close to 1.0 Gyr, with
good consistency between the three sets of models.  The Dartmouth models
imply a somewhat younger age, but that group does not currently
tabulate isochrones for ages less than 1.0 Gyr. The magnitude and
color shifts needed to force agreement with the secondary star range
from $(m-M)_V = 10.27$ to 10.34 and $E(B-V) = 0.05$ to 0.08, where the
PARSEC, Victoria-Regina, and Yonsei-Yale model values agree well and
sit at the low end for both quantities.

Systematic uncertainties are always important considerations for age
measurements, and we briefly discuss them here. Metallicity
uncertainties, whether due to the cluster or to solar composition,
affect the ages at the level of a few hundredths of a Gyr (see the
PARSEC and Victoria-Regina panels of Figs. \ref{tomr} and
\ref{cmdfix}). There are a number of stellar physics uncertainties
that could affect whether the isochrones truly represent observed
stars near the turnoff. One of these is rotation
\citep[e.g.,][]{rotmods}, which is expected to extend the lifetimes of
main sequence stars and make them appear more luminous. While the
stars of the eclipsing binary should not be affected, ages determined
from rotating turnoff stars would be biased toward younger ages. This
effect is difficult to evaluate properly, and may require examination
of spectroscopic rotational velocities for a large sample of the turnoff
stars.

For most of our comparisons utilizing the components of KIC 9777062, we will
be insensitive to the details of convective core overshooting because
this has its most observable effects near core hydrogen exhaustion or
beyond. So at present, the photometry of the bright main sequence
stars is our main constraint on this process. Without a more detailed
survey of the radial velocities of turnoff stars to evaluate binarity,
we cannot say much more than the cluster CMD seems to be consistent
with overshooting in an amount of approximately 0.2 pressure scale
heights, as is used in the Dartmouth, Yonsei-Yale, and PARSEC models.

\subsubsection{Age Summary}

Because we produced measurements covering a
disparate set of techniques, it is worth summarizing the
results and examining the degree to which the different
techniques produce consistent results.

Asteroseismic masses and radii for clump giants in the cluster imply
an age near 0.9 Gyr, but this estimate
may suffer from significant systematic uncertainties, including the
treatment of convection in the isochrones and the asteroseismic
scaling relations used to calculate the radii.

The masses and radii measured for the stars in the eclipsing binary
imply different ages, with the primary appearing younger than the
secondary star ($1.02 - 1.08$ Gyr versus $1.17 - 1.25$ Gyr). Because
the primary star shows the Am metal-line phenomenon, we believe it is
more likely to be experiencing physics effects (like diffusion) that
are not properly represented in the models used to interpret the
age. If the two stars in the binary have lower metallicity than used
in the model isochrones, this would also tend to slightly improve the
agreement of the ages derived separately from each of them.

Age determinations from the CMD require some decision on how much the
isochrones need to be shifted in magnitude and color. While fits are
frequently done to derive distance, reddening, and age simultaneously,
uncertainties in the color-$\teff$ relations assumed in the isochrones
affect the validity of the inferred values. If we shift isochrones to
match the $V$ magnitude of the secondary star in the eclipsing binary
at its measured mass, the implied distance modulus is in rough
agreement with our independently derived values from fitting a
period-luminosity relation to the cluster $\delta$ Sct stars and from
asteroseismology of the giant stars.  However, the preferred CMD age in
this case appears to be near 1.0 Gyr (uncertainty around 0.05 Gyr), in
disagreement with the secondary star mass-radius measurement.  It
should be remembered that the CMD age is dependent on our ability to isolate
a sample of single stars that are brighter than the turnoff ($V \la
12$). In a sparse cluster like NGC 6811, this is challenging, and more
radial velocity and asteroseismic work can be done to verify that the
stars in our sample do not have bound companions. In addition, if
cluster stars have higher metallicity than used in the isochrones, the
CMD age would move older. A higher metallicity makes
the mass-radius ages of the KIC 9777062 stars disagree more, however,
and leads to the conclusion that consistency in the ages cannot be
achieved by simply assuming a different metal abundance.

\section{Conclusions}

We presented a detailed analysis of the first detached eclipsing
binary member of the open cluster NGC 6811 containing stars near the
turnoff. The primary star shows surface abundance anomalies consistent
with being a metal-line Am star, and one of the stars (probably the
secondary, based on its CMD position, radial velocity scatter, and
increased photometric scatter during primary eclipse) is a $\gamma$
Dor pulsating star. The indication of a convective blueshift for
  the primary star implies that it probably has a turbulent surface
  convection zone that is shallow enough to allow diffusive chemical
  separation and the Am phenomenon. The less massive secondary star
  does not show chemical peculiarities and shows a lower
  microturbulent velocity, implying that its surface convection zone
  goes deep enough to circumvent diffusive separation. Nearly all
  stars in well-measured binaries with masses between 1.3 and $1.55
  \msun$ appear chemically normal.

A surprising outcome of the measurement of the
stellar masses and radii is that they indicate different ages, with
the secondary star returning an age that is 15-20\% larger than the
primary star. In lower mass and shorter period binaries this kind of
discrepancy has been attributed to induced magnetic activity and
inhibition of convective energy transport that leads to a larger
radius for the secondary star having a deeper convective zone. While
the stars are rotating slowly and thereby show signs of tidal
interaction, the orbital period is much longer than any binaries
showing radius discrepancies among lower mass stars and the convection
zones in these stars would themselves be much lower in mass and less
likely to be affected by magnetic activity. On the other hand, more than
a third of Am stars in the fields observed by \kep have been shown to
have photometric variations that are likely due to spots
\citep{balona15}.
Further, there are other well-studied eclipsing
binaries containing more massive Am stars that show similar patterns
in their masses and radii. Regardless, we have yet to identify a convincing
explanation why at least one of these stars seems to have an anomalous radius when compared with models.

Deeper study of other types of stars in the cluster may help clarify
the picture.  With the large number of $\delta$ Sct pulsating stars
occupying the turnoff of this cluster, this binary and others provide
the excellent opportunity to determine accurate masses for an ensemble
of pulsating stars with extensive \kep photometry in order to derive
additional strong observational constraints on stellar models. One
application we examined was the use of the period-luminosity relation
for high-amplitude $\delta$ Sct stars with {\it Hipparcos} parallaxes
to derive an independent measure of the distance modulus. A promising
line of future research is the derivation of mean stellar densities
from the equivalent of the large frequency spacing for these stars
\citep{suarez}.

We re-examined the sample of secondary clump and asymptotic giant stars in
the cluster and re-analyzed their asteroseismic data from the main
\kep field. The small sample of clump stars (there are no first-ascent
red giants) and the relatively large measurement uncertainties do not
allow us to produce a tight constraint on the age, but the mass-radius
values are consistent with a 0.9 Gyr age.
A deeper analysis going beyond the observables $\Delta \nu$ and
$\nu_{\rm max}$ could further test the consistency of the asteroseismic
results with measurements from the CMD, the main sequence pulsators, and
the eclipsing binary stars.

At the end of this survey, we have not found a model of any age that
is consistent with all of the observations. The results tend to break
in to groups near 1.0 Gyr (the mass and radius of the primary of the
eclipsing binary; photometry of turnoff stars when isochrones are
pinned to the mass and photometry of the secondary star; the
asteroseismic masses and radii of the secondary clump stars) or near
1.2 Gyr (the mass and radius of the secondary star). The distance
  moduli we have derived from the eclipsing binary and $\delta$ Sct
  stars also differ with each other at the 0.1 mag level, with the
  $\delta$ Sct result being closer to the isochrone predictions.
There are significant differences between the predictions of different
sets of isochrones and significant uncertainties that currently limit
our ability to identify a preferred age and distance.  Observations that would
improve the situation include: further radial velocity study of the
brightest main sequence stars (especially the subgiant candidate KIC
9777532) to find or rule out previously undetected binary companions;
characterization of additional eclipsing binaries in the cluster; and
establishment of more precise spectroscopic temperatures for turnoff
stars. On the theoretical side, there is a need to identify the
sources of the differences between sets of isochrones, and to test and
improve the treatment of helium and heavy element diffusion. At the
turnoff of NGC 6811, the mix of stars --- with and without binary
companions, and with different rates of rotation --- may be
confounding our understanding more than we realize.

\acknowledgments We thank the anonymous referee for comments that
improved the presentation of the paper.

E.L.S. is grateful to the Stellar Astrophysics Centre
at Aarhus University for their generosity and hospitality during his
sabbatical stay, during which part of this work was completed. Funding
for the Stellar Astrophysics Center is provided by the Danish National
Research Foundation (grant agreement no. NDRF106) with research
supported by the ASTERISK project (ASTERoseismic Investigations with
SONG and Kepler) funded by the European Research Council (grant
agreement no. 267864). Our work has been funded through grant AST
09-08536 from the National Science Foundation and grant NNX13AC19G
from the National Aeronautics and Space Administration to E.L.S.
M.L. was supported as part of the Research Experiences for
Undergraduates site at San Diego State University, funded by the
National Science Foundation under grant AST-0850564 to E.L.S.
K.B. acknowledges support from the Carlsberg Foundation and the Villum
Foundation. We would also like to thank S. Brunker, E. Rich,
C. Curtin, M. Lapid, J. Mascoop, and J. Pautzke for assisting in the
acquistion of ground-based photometric observations, and staff and
student support astronomers at the NOT (T. Augusteijn, A. A. Djupvik,
T. Pursimo, J. Telting, F. S. Kiaeerad, G. Barisevicius, J. Lehtinen,
O. Smirnova, S. Geier, T. Kangas, J. Kajava, and Y. Martinez-Osori) for
assisting in the acquisition of FIES observations.

We are very grateful to the \kep team for the opportunity to work with
such a precise and extensive dataset for detecting variable
stars. This paper includes data collected by the \kep mission. Funding
for the \kep mission is provided by the NASA Science Mission
directorate. This research made use of the SIMBAD database, operated
at CDS, Strasbourg, France; the NASA/ IPAC Infrared Science Archive,
which is operated by the Jet Propulsion Laboratory, California
Institute of Technology, under contract with the National Aeronautics
and Space Administration; the WEBDA database, operated at the
Institute for Astronomy of the University of Vienna; Astropy, a
community-developed core Python package for astronomy (Astropy
Collaboration, 2013); and the Mikulski Archive for Space Telescopes
(MAST). STScI is operated by the Association of Universities for
Research in Astronomy, Inc., under NASA contract NAS5-26555. Support
for MAST is provided by the NASA Office of Space Science via grant
NNX09AF08G and by other grants and contracts.

The Hobby–Eberly Telescope (HET) is a joint project of the University
of Texas at Austin, the Pennsylvania State University, Stanford
University, Ludwig-Maximilians-Universit\"{a}t M\"{u}nchen, and
Georg-August-Universit\"{a}t G\"{o}ttingen. The HET is named in honor of its
principal benefactors, William P. Hobby and Robert E. Eberly.

Funding for SDSS-III has been provided by the Alfred P. Sloan Foundation, the
Participating Institutions, the National Science Foundation, and the
U.S. Department of Energy Office of Science. The SDSS-III web site is
{\tt http://www.sdss3.org/}. SDSS-III is managed by the Astrophysical Research Consortium for the
Participating Institutions of the SDSS-III Collaboration including the
University of Arizona, the Brazilian Participation Group, Brookhaven National
Laboratory, Carnegie Mellon University, University of Florida, the French
Participation Group, the German Participation Group, Harvard University, the
Instituto de Astrofisica de Canarias, the Michigan State/Notre Dame/JINA
Participation Group, Johns Hopkins University, Lawrence Berkeley National
Laboratory, Max Planck Institute for Astrophysics, Max Planck Institute for
Extraterrestrial Physics, New Mexico State University, New York University,
Ohio State University, Pennsylvania State University, University of
Portsmouth, Princeton University, the Spanish Participation Group, University
of Tokyo, University of Utah, Vanderbilt University, University of Virginia,
University of Washington, and Yale University.

\appendix

\section{Photometric Calibration}

The field of the eclipsing binary KIC 9777062 was observed on 2012 Aug. 26, while three other fields
(including the cluster center) were observed on 2013 Aug. 8. We
observed Landolt standard fields SA 92, SA 110, and SA 114 as well as
the open clusters NGC 6939 and NGC 6940 at airmasses ranging from
1.030 to 1.846. We used stars and standard values from \citet[Nov. 5,
  2013 release]{stetson} covering a color range $0 \le (B-I) \le
8$. In the course of the calibration, we discovered that the residuals of our
photometry of the NGC 6940 field had a consistent trend with $x$
position. Upon investigation, we found that this trend (a large 0.09
mag in full extent) appears to be in the Stetson standard magnitudes
because we find it consistently in these two nights and others. We
opted to use our own observations to recalibrate the NGC 6940 stars
because we observed this cluster over a significantly larger range of
airmass than our other standard fields and would therefore be in a
better position to evaluate airmass-dependent terms in the photometric
transformations below.

We derived instrumental magnitudes from aperture photometry conducted using
DAOPHOT with curve-of-growth corrections derived using DAOGROW.
The transformations to the standard system employed the
equations
\begin{eqnarray} 
b &=& B + a_0 + a_1(B-I) + a_2X, \nonumber \\
v &=& V + b_0 + b_1(B-I) + b_2X, \nonumber \\
i &=& I + c_0 + c_1(B-I) + c_2(B-I)^2 + c_3X, \nonumber
\end{eqnarray}
where $b$, $v$, and $i$ are instrumental magnitudes, $B$, $V$, and $I$ are
standard system magnitudes, and $X$ is airmass. 

\begin{figure*}
\includegraphics[scale=0.6]{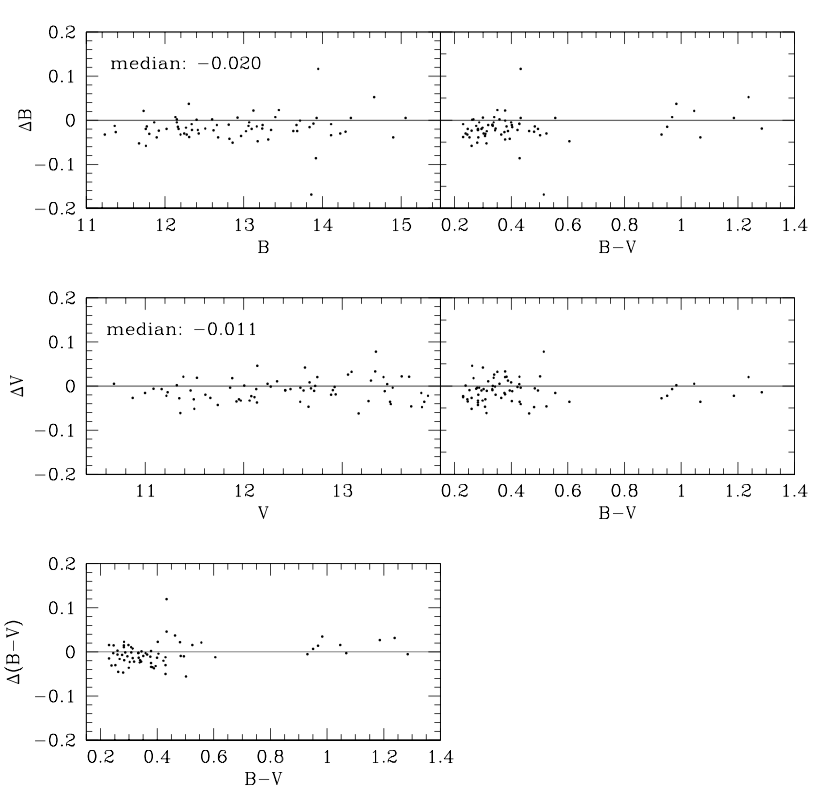}
\caption{Differences between the $BV$ photometry of \citet{glush} and
  the present paper in the sense of (ours - theirs).\label{glush}}
\end{figure*}

We compared our $BV$ photometry with two previously published
datasets, partly to provide an independent check of literature
photometry and to test whether different photometric observations of
this extended cluster could realistically be combined together. Using
82 stars in common with the photoelectric photometry of \citet{glush},
we found median offsets (in the sense of our photometry minus theirs)
of $-0.020$ and $-0.011$ mag in $B$ and $V$, respectively.  There were
no signs of magnitude dependencies in the differences (see
Fig. \ref{glush}), and only a small sign of a trend in color residuals
versus color (in the sense that giants tended to be slightly redder
relative to the main sequence stars in our photometry).

\begin{figure*}
\includegraphics[scale=0.4]{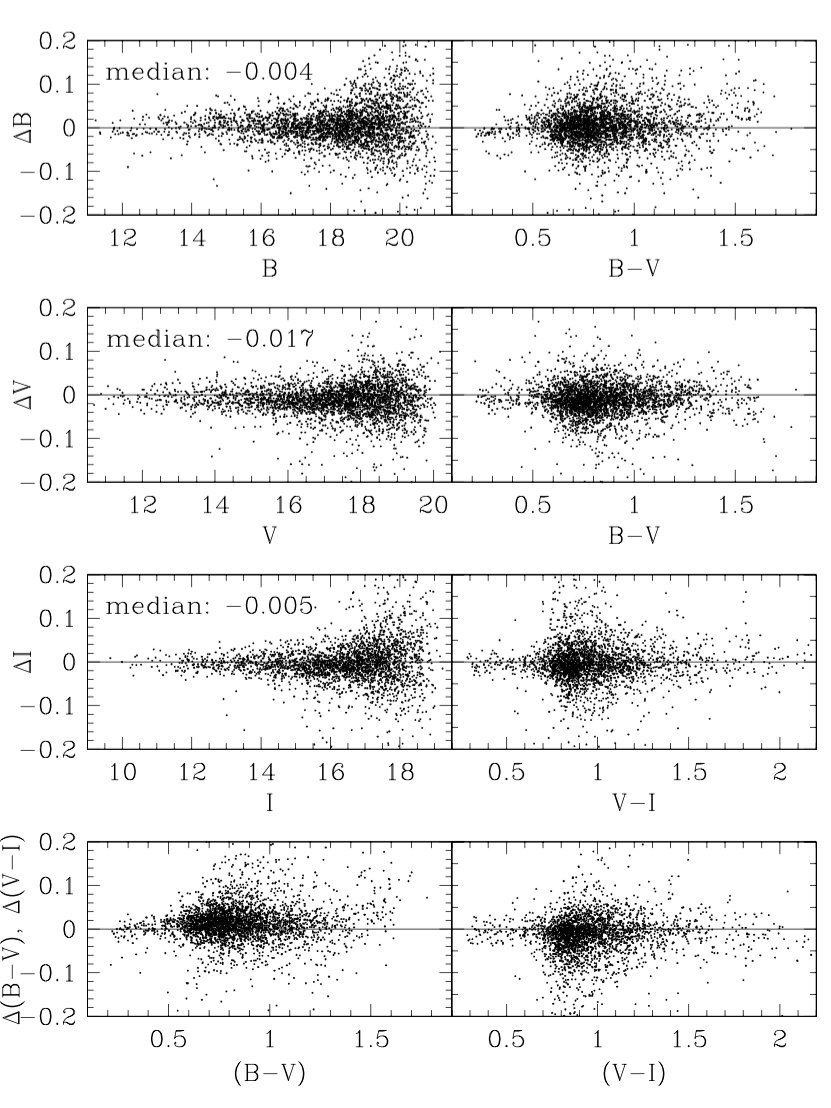}
\caption{Differences between the $BVI_C$ photometry of \citet{janes} and the present paper in the sense of (ours - theirs).\label{rsdjanes}}
\end{figure*}

\begin{figure*}
\includegraphics[scale=0.6]{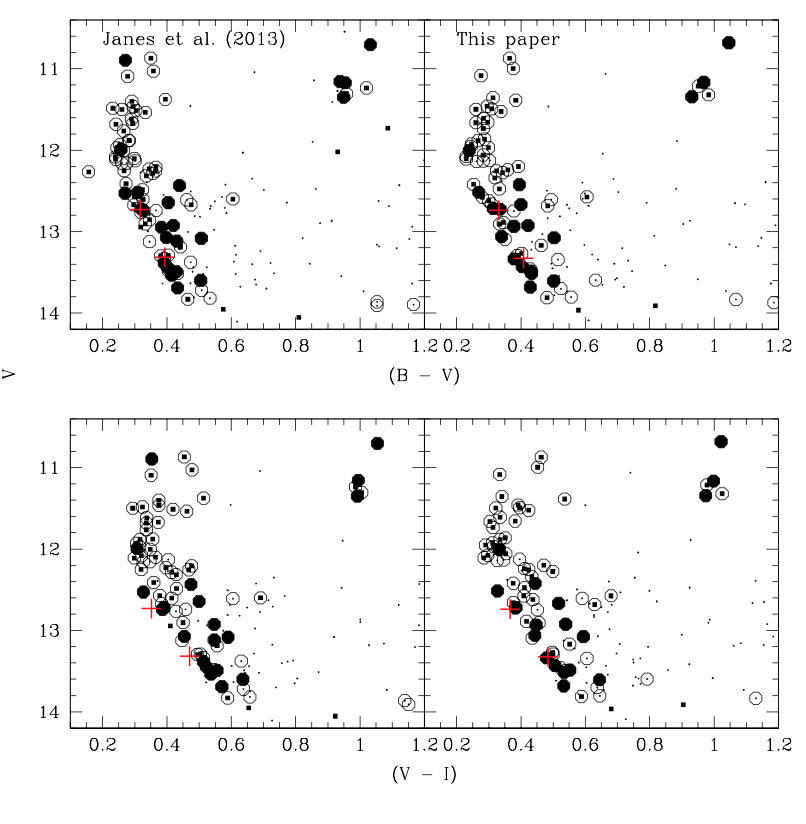}
\caption{Color-magnitude diagrams for stars in the NGC 6811 field in
  $BV$ (top row) and $VI_C$ (bottom row) from \citet{janes} and this
  work. Stars with proper motion memberships $P_\mu > 50$\% in any of the
  three studies \citep{sanders6811,dias,khar13} are shown with circles,
  and if proper motion membership is corroborated by radial velocities
  \citep{frinch,merm,mz,meibom11} the circle is filled. Known or
  candidate binaries are indicated with small square symbols. The
  components of KIC 9777062 are shown with $+$.  \label{cmdmem}}
\end{figure*}

We also compared with the photometry of \citet{janes} and found good
agreement (see Fig. \ref{rsdjanes}). There were no indications of
trends with magnitude, hints of a small trend of $B$ magnitude with
color, and slight trends of magnitude residuals with declination
coordinate (most significantly, a variation of about 0.04 mag in $B$
in the southernmost of our four fields, in the sense that our
photometry gets systematically fainter with respect to
\citeauthor{janes} as declination decreases). In their study and ours, the
cluster was covered by tiling together multiple fields, but
\citeauthor{janes} also made use of multiple telescope/instrument
combinations. The section of the field that shows the greatest
evidence of a position-dependent trend was their Hall telescope field
from 2010. Based on this, we believe we have done a slightly better
job of removing position-dependent trends, but the \citeauthor{janes} study
has somewhat better signal-to-noise, as seen in reduced scatter among main sequence stars in Figure \ref{cmdmem}.

\section{Masses and Radii of Am/Fm Stars}

\begin{figure*}
\includegraphics[scale=0.5]{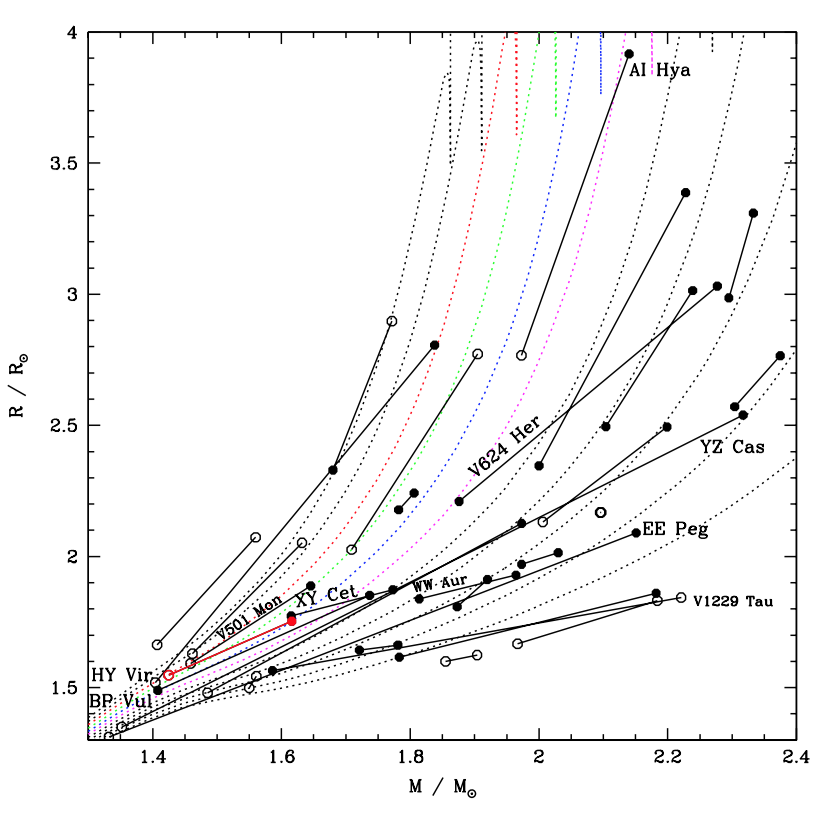}
\caption{Mass-radius plot for the members of well-studied eclipsing
  binary stars. Am/Fm stars are shown with solid points, and members
  of the same binary are connected with line segments. KIC 9777062 is
  shown in red. PARSEC isochrones with $Z=0.0137$ and ages between 0.5
  and 1.4 Gyr (0.1 Gyr spacing) are shown.
\label{amstars}}
\end{figure*}

We looked at systems in the compilation of \citet{torres} that
had stars in the mass range covered by these stars (approximately
$1.4-2.4 \msun$), supplemented with additional eclipsing Am binaries
from more recent literature
\citep{groene,south11,torres12,torres15}. Detached systems like these
should contain stars with equal age and bulk composition (although the
surface composition may be greatly affected), so that each pair {\it
  should} lie on the same isochrone if the Am/Fm phenomenon is
unimportant. In Fig. \ref{amstars}, we plot a $M-R$ diagram for these
stars using isochrones of NGC 6811 metal content. While not
universally true (for stars near the masses of the KIC 9777062
components, see HY Vir and V501 Mon), there are a number of systems
with a line connecting the components that appears to have a shallower
slope than the isochrones. In some of these cases (YZ Cas,
\citealt{pavlov}; EE Peg, \citealt{eepeg}; V1229 Tau,
\citealt{groene}), the slope can be attributed to a low
age\footnote{For V1229 Tau, the system is a member of the Pleiades and
  can therefore be dated to near 100 Myr. For YZ Cas, the secondary
  star is of F2 spectral type, and does not show the metal-line
  phenomenon. Its photospheric composition
  can therefore constrain the bulk composition of the stars to be near
  solar.}.
However, in other cases it is less clear that age or composition can
be used to explain the system properties. Among the more important
examples:
\begin{itemize}
\item BP Vul \citep{lacy}: Both stars in this system show the
  metal-line phenomenon, and so the bulk composition cannot be
  inferred from the system itself. However, the stars are close
  to those of KIC 9777062 in $M-R$ space, and they share a very
  similar slope in $M-R$ plane. The
  authors found that the primary star implied a slightly younger age
  than the secondary star for a common bulk composition. With the
  caveat that the bulk composition is not well determined, the
  inferred age of the system is also quite similar to that of KIC
  9777062.  It should be noted that secondary of BP Vul is one of the
  least massive stars in this binary star sample showing the Fm
  phenomenon --- no other binary component in the range $1.4 \le M /
  \msun \le 1.58$ is known to show metal line anomalies.
\item XY Cet \citep{south11}: This system is composed of two Am stars
  in a short period (2.78 d) binary, and the secondary star has
  a mass and radius that are within about 1\% of those of the primary of KIC
  9777062. The bulk composition of the stars
  cannot be inferred from spectroscopy, but the authors found that the
  $M-R$ line for the binary was less steep than predicted by any
  models they used. The best match was to solar abundance models with
  age near 850 Myr.
\item WW Aur \citep{south05}: The line segment connecting the stars in
  this system presents a rather shallow slope that undoubtedly helped
  push the authors to infer a {\it highly} super-solar metal abundance and
  young age (90 Myr). It should be noted that those authors did not
  see evidence that the Am phenomenon was the cause of the unusual
  stellar properties.
\item Other systems showing shallower slopes in the $M-R$ plane than
  expected: AI Hya \citep{aihya}, V624 Her \citep{v624her}, HW CMa \citep{torres12}
\end{itemize}
Binaries that show this radius discrepancy also appear to show a
discrepancy in the relative magnitudes of the two stars, in the sense
that the primary star tends to be fainter than expected based on the
mass and brightness of the secondary star. KIC 9777062 follows this
pattern as well (see \S \ref{cmd}).

KIC 9777062 is a valuable addition to the list above because there are
constraints on the bulk composition of the stars from both the
secondary star and from other stars in the cluster itself. Diffusion
is important in the interiors of both stars given that they are both
rotating slowly and are massive enough to have mostly radiative
envelopes. However, only the primary shows surface heavy element
abundances consistent with being a metal-lined star. The lack of
$\delta$ Sct pulsation from the primary may indicate that helium is
draining out of the He II ionization zone that drives the pulsation
--- a significant fraction of Am stars pulsate, but usually at lower
amplitude levels than typical $\delta$ Sct stars
\citep{smalley}. Models, however, predict that diffusion should result
in larger radii \citep{turcotte} due to increased metal opacity in the
surface layers, which runs contrary to what we see in the KIC 9777062
binary. If the Am stars in KIC 9777062 and other binaries are indeed
anomalously small in radius, the physical mechanism is still unclear
and it is odd that the effect only shows up in some binaries.

\clearpage
\begin{turnpage}
\begin{deluxetable*}{lccccccccc}
\tablewidth{0pc}
\tabletypesize{\scriptsize}
\tablecaption{Giant Stars in the NGC 6811 Field}
\tablehead{\multicolumn{10}{l}{Identifications\tablenotemark{a}:}\\
  \colhead{KIC} & \colhead{9409513} & \colhead{9716522} & \colhead{9776739} & \colhead{9532903} & \colhead{9655101} & \colhead{9534041} & \colhead{9655167} & \colhead{9716090} & \colhead{9897838}\\
\colhead{S71/G99} & \colhead{G53} & \colhead{170} & \colhead{140} & \colhead{194} & \colhead{95} & \colhead{G47} & \colhead{106} & \colhead{92} & \colhead{G59}\\  
\colhead{J13} & \colhead{} & \colhead{5441} & \colhead{4726} & \colhead{} & \colhead{3278} & \colhead{} & \colhead{3603} & \colhead{3202} & \colhead{}\\
\colhead{Tycho\tablenotemark{b}} & \colhead{815} & \colhead{2634} & \colhead{118} & \colhead{103} & \colhead{530} & \colhead{3683} & \colhead{} & \colhead{2356} & \colhead{3343}}
\startdata
$\alpha$ (2000) & 19:36:04.25 & 19:37:34.62\tablenotemark{c} & 19:37:22.07 & 19:37:50.18 & 19:36:57.13 & 19:39:32.94 & 19:37:02.68 & 19:36:55.81 & 19:38:31.12 \\ 
$\delta$ (2000) & +45:58:27.5 & +46:24:09.9\tablenotemark{c} & +46:32:50.6 & +46:07:46.5 & +46:22:42.6 & +46:11:55.4 & +46:23:13.1 & +46:27:37.7 & +46:47:33.3 \\ 
\hline
\multicolumn{10}{l}{Photometry:} \\
$V$ & $10.543\pm0.033$ & $10.703\pm0.018$ & $11.157\pm0.015$ & $11.165\pm0.002$ & $11.235\pm0.016$ & $11.272\pm0.016$ & $11.306\pm0.003$ & $11.350\pm0.006$ & $11.142\pm0.001$ \\ 
$B-V$ & $0.952\pm0.007$ & $1.032\pm0.009$ & $0.938\pm0.007$ & $0.968\pm0.003$ & $1.020\pm0.083$ & $0.930\pm0.019$ & $0.958\pm0.001$ & $0.949\pm0.016$ & $0.857\pm0.025$ \\
$V-I$ & & $1.054\pm0.005$ & $0.995\pm0.002$ & $0.998\pm0.004$ & $0.986\pm0.002$ & & $1.005\pm0.002$ & $0.992\pm0.002$ & \\
$V-K_s$ & $2.297\pm0.035$ & $2.440\pm0.023$ & $2.23\pm0.019$ & $2.264\pm0.012$ & $2.241\pm0.019$ & $2.202\pm0.021$ & $2.287\pm0.011$ & $2.264\pm0.013$ & $2.191\pm0.011$\\
\hline
$T_{\rm eff}$\tablenotemark{d} (K) & 4950 & 4805 & 5025 & 4990 & 5010 & 5060 & 4960 & 4990 & \\
$T_{\rm eff}$\tablenotemark{e} (K) & 5065 & 4952 &      & 5105 & 5111 & 5107 & 5091 &      & \\
$T_{\rm eff}$\tablenotemark{f} (K) &      &      & 4952 & 5008 & 5005 &      & 4924 & 4980 & \\
\mbox{[Fe/H]}\tablenotemark{e} & $0.10\pm0.03$ & $0.04\pm0.03$ & & $0.02\pm0.03$ & $-0.03\pm0.03$ & $-0.03\pm0.03$ & $-0.01\pm0.03$ & & \\
$\Delta \nu$ ($\mu$Hz) & $6.044\pm0.026$ & $4.856\pm0.019$ & $7.621\pm0.032$ & $7.548\pm0.029$ & $7.871\pm0.023$ & $8.375\pm0.026$ & $8.040\pm0.033$ & $8.519\pm0.035$ & \\
$\nu_{\rm max}$ ($\mu$Hz) & $70.69\pm0.97$ & $54.10\pm1.08$ & $94.86\pm2.29$ & $93.15\pm1.81$ & $98.71\pm2.64$ & $108.94\pm1.85$ & $105.33\pm5.97$ & $109.68\pm2.47$ & \\
$M$ ($\msun$) & $2.29\pm0.12$ & $2.35\pm0.16$ & $2.24\pm0.18$ & $2.18\pm0.15$ & $2.21\pm0.19$ & $2.35\pm0.14$ & $2.43\pm0.42$ & $2.19\pm0.17$ & \\
$R$ ($\rsun$) & $10.45\pm0.20$ & $12.20\pm0.29$ & $8.88\pm0.24$ & $8.86\pm0.21$ & $8.65\pm0.25$ & $8.47\pm0.18$ & $8.81\pm0.51$ & $8.19\pm0.21$ & \\
\hline
\multicolumn{10}{l}{Membership:} \\
S71       &    & 97 & 97 & 97 & 97 &    & 97 & 94 & \\
UCAC4     & 36 & 99 & 99 & 99 & 99 & 99 & 99 & 99 & 0 \\
K13       & 27 & 93 & 87 & 83 & 17 & 13 & 22 & 95 & 0 \\
RV        & Y  & Y  & Y  & Y  & B  & Y  & B  & Y  & Y \\
Seismic   & ?  & Y  & Y  & Y  & Y  & Y  & Y  & Y  & N\\
Position  & N  & Y  & Y  & N  & Y  & N  & Y  & Y  & N\\
\hline
Notes: & AGB? binary? & AGB & RC & RC & RC, SB1? & RC? & RC, SB1 & RC & nonmem.
\enddata
\tablenotetext{a}{S71: \citet{sanders6811}; J13: \citet{janes}; G99: \citet{glush}; UCAC4: \citet{ucac4}, \citet{dias14}; K13: \citet{khar13}.}
\tablenotetext{b}{Full {\it Tycho} ID is TYC 3556-XXXXX-1.}
\tablenotetext{c}{Incorrect position given in SIMBAD and \citet{merm08}.}
\tablenotetext{d}{from $V-K_s$ color using \citet{rm05} relations with $E(B-V)=0.07$.}
\tablenotetext{e}{from APOGEE (Data Release 12).}
\tablenotetext{f}{from \citet{mz}.}
\label{giants}
\end{deluxetable*}
\end{turnpage}

\begin{turnpage}
\begin{deluxetable*}{rrrcccccccccccl}
\tablewidth{0pc}
\tabletypesize{\scriptsize}
\tablecaption{Pulsating Main Sequence Stars in the NGC 6811 Field}
\tablehead{\multicolumn{4}{c}{Identifications\tablenotemark{a}} & \multicolumn{2}{c}{Position}
  & \multicolumn{3}{c}{Photometry} & \multicolumn{5}{c}{Membership\tablenotemark{a}} & 
  \colhead{Notes\tablenotemark{c}} \\
\colhead{S71} & \colhead{J13} & \colhead{Tycho\tablenotemark{b}} & \colhead{KIC} & \colhead{$\alpha$ (2000)} & \colhead{$\delta$ (2000)} & \colhead{$V$} & \colhead{$B-V$} & \colhead{$V-I$} & \colhead{S71} & \colhead{UCAC4} & \colhead{K13} & \colhead{RV} & \colhead{Pos} & }
\startdata
108 & 3636 & 1668 & 9655177 & 19:37:03.23 & +46:19:25.7 & $10.869\pm0.012$ & $0.350\pm0.016$ & $0.454\pm0.002$ & 96 & 99 & 78 & B & Y & $\delta$ Sct (vC,L,U,D); 1\\
205 & 6724 & 1676 & 9594857 & 19:37:58.76 & +46:14:19.4 & $11.028\pm0.014$ & $0.358\pm0.016$ & $0.478\pm0.002$ & 89 & 99 & 87 & B & N & $\delta$ Sct (vC,L,D); 2\\
136\tablenotemark{d} & 4690 & 1768 & 9716385 & 19:37:21.48 & +46:24:33.8 & $11.092\pm0.004$ & $0.278\pm0.002$ & $0.351\pm0.002$ & 96 & 99 & 87 & B & Y & $\delta$ Sct (vC,L); 3\\
 86 & 2988 & 1838 & 9655055 & 19:36:51.91 & +46:23:20.4 & $11.400\pm0.011$ & $0.291\pm0.007$ & $0.375\pm0.211$ & 97 & 99 & 66 & B & Y & $\delta$ Sct (D,U)\\
 75 &      & 2456 & 9532168 & 19:36:47.21 & +46:09:48.8 & $11.41$          & 0.24            &                 & 60 & 99 & 90 & B & N & $\delta$ Sct (B), 3\\
166 & 5314 & 1698 & 9655514 & 19:37:32.10 & +46:19:15.0 & $11.464\pm0.013$ & $0.295\pm0.006$ & $0.375\pm0.003$ & 93 & 99 & 19\tablenotemark{e} & B & Y & $\delta$ Sct (vC,L,U), 4\\
87 & 3004 & 2336 & 9776378 & 19:36:52.01 & +46:32:05.1 & $11.484\pm0.003$ & $0.232\pm0.023$ & $0.324\pm0.003$ & 97 & 99 & 81 & B & Y & $\delta$ Sct (B), 5\\
144 & 4845 &  716 & 9655422 & 19:37:24.09 & +46:23:52.1 & $11.512\pm0.014$ & $0.304\pm0.008$ & $0.419\pm0.003$ & 54 & 99 & 98 & B & Y & $\delta$ Sct (vC,L,U,D)\\
121 & 4249 & 2492 & 9716301 & 19:37:13.77 & +46:25:25.7 & $11.618\pm0.007$ & $0.290\pm0.009$ & $0.337\pm0.003$ & 96 & 99 & 52 & N & Y & B variable (B)\\
149 & 4946 &      & 9655444 & 19:37:25.64 & +46:18:36.8 & $11.671\pm0.007$ & $0.294\pm0.008$ & $0.373\pm0.003$ & 95 & 99 &  0 & B & Y & $\delta$ Sct (D)\\
127 & 4453 &  856 & 9655346 & 19:37:17.10 & +46:23:14.7 & $11.763\pm0.007$ & $0.266\pm0.007$ & $0.336\pm0.003$ & 93 & 99 & 91 & B & Y & EW with $\delta$ Sct; 6\\
    & 9909 & 3387 & 9595743 & 19:39:19.22 & +46:14:57.5 & $11.812\pm0.008$ & $0.319\pm0.005$ & $0.393\pm0.003$ &    &  0 &  0 &   & N & 7\\
113 & 3765 & 1344 & 9716220 & 19:37:05.46 & +46:24:58.5 & $11.879\pm0.007$ & $0.282\pm0.005$ & $0.356\pm0.003$ & 95 & 99 & 63 & B & Y & $\delta$ Sct (L,D,B)\\
230 & 7342 & 1690 & 9777351 & 19:38:16.13 & +46:31:31.9 & $11.930\pm0.008$ & $0.254\pm0.002$ & $0.306\pm0.040$ & 97 & 99 & 97 & B & N & hybrid; 8\\
157 & 5067 & 1888 & 9655470 & 19:37:27.79 & +46:23:10.1 & $12.003\pm0.009$ & $0.265\pm0.013$ & $0.348\pm0.002$ & 96 & 99 & 72 & B & Y & $\delta$ Sct (L,D)\\
192 & 6195 &  914 & 9716667 & 19:37:48.08 & +46:27:25.3 & $12.082\pm0.011$ & $0.240\pm0.003$ & $0.321\pm0.008$ & 92 & 99 & 99 & B & Y & $\delta$ Sct (D); 9\\
97  & 3339 &  944 & 9655114 & 19:36:58.21 & +46:20:23.9 & $12.101\pm0.024$ & $0.299\pm0.028$ & $0.365\pm0.003$ & 97 & 99 & 86 & B & Y & $\delta$ Sct (vC,L,U,B); 10\\
206 & 6846 & 2232 & 9655909 & 19:38:01.74 & +46:19:07.0 & $12.131\pm0.004$ & $0.301\pm0.001$ & $0.404\pm0.003$ & 94 & 99 & 93 &   & Y & $\delta$ Sct (D)\\
146 & 4896 &      & 9655433 & 19:37:24.86 & +46:18:39.0 & $12.159\pm0.005$ & $0.260\pm0.007$ & $0.326\pm0.003$ & 96 & 87 &  0 &   & Y & $\delta$ Sct (B); 11\\
219 & 7013 &  382 & 9837659 & 19:38:06.73 & +46:36:30.7 & $12.208\pm0.005$ & $0.366\pm0.007$ & $0.477\pm0.006$ &  0 & 12 & 61 & B & N & $\delta$ Sct (D); 10\\
195 & 6335 &  370 & 9777062 & 19:37:50.59 & +46:35:23.2 & $12.230\pm0.008$ & $0.345\pm0.011$ & $0.397\pm0.004$ & 93 & 99 & 30 & Y & N & $\gamma$ Dor; 12\\
 81 & 2852 & 2038 & 9594022 & 19:36:49.68 & +46:14:26.3 & $12.252\pm0.006$ & $0.267\pm0.014$ & $0.320\pm0.004$ & 89 & 99 &  0 & B & Y & $\delta$ Sct (D,B)\\
147 & 4918 &      & 9655438 & 19:37:25.23 & +46:19:35.7 & $12.255\pm0.006$ & $0.365\pm0.003$ & $0.468\pm0.003$ & 97 & 99 & 91 & B & Y & hybrid (U); 8, 10\\
115 & 3998 &      & 9716256 & 19:37:09.67 & +46:27:09.0 & $12.273\pm0.013$ & $0.319\pm0.003$ & $0.403\pm0.002$ &  0 &  0 & 0 & Y & Y & $\gamma$ Dor; 8\\
138 & 4735 & 1762 & 9655402 & 19:37:22.18 & +46:18:51.2 & $12.286\pm0.041$ & $0.356\pm0.051$ & $0.415\pm0.003$ &  0 & 73 & 46 & B & Y & $\delta$ Sct (D); 10\\
165 & 5258 & 1248 & 9655501 & 19:37:31.22 & +46:21:32.0 & $12.411\pm0.007$ & $0.273\pm0.010$ & $0.359\pm0.003$ & 93 & 99 & 43 & B & Y & $\delta$ Sct (D), hybrid (U)\\
143 & 4818 &      & 9655419 & 19:37:23.63 & +46:23:26.7 & $12.486\pm0.004$ & $0.324\pm0.005$ & $0.430\pm0.003$ & 90 & 99 & 50 & B & Y & $\delta$ Sct (L,D)\\
169 & 5377 &  306 & 9837267 & 19:37:33.59 & +46:37:08.8 & $12.528\pm0.011$ & $0.270\pm0.019$ & $0.327\pm0.018$ & 95 & 99 & 30 & B & N & $\delta$ Sct (D)\\
255 & 8365 & 3564 & 9533483 & 19:38:41.15 & +46:07:01.4 & $12.549\pm0.009$ & $0.390\pm0.012$ &                 &  0 &  0 &  0 & N & N & SPB (D)\\
135 & 4687 &      & 9655393 & 19:37:21.37 & +46:19:53.3 & $12.571\pm0.005$ & $0.299\pm0.008$ & $0.378\pm0.003$ & 94 & 99 & 84 & B & Y & $\delta$ Sct (L,U,B); 8\\
201 & 6501 &      & 9655800 & 19:37:53.16 & +46:18:28.8 & $12.644\pm0.010$ & $0.404\pm0.009$ & $0.500\pm0.004$ & 94 & 99 & 70 & Y & Y & $\gamma$ Dor (U,D)\\
119 & 4182 &      & 9655288 & 19:37:12.44 & +46:23:29.3 & $12.672\pm0.005$ & $0.297\pm0.008$ & $0.384\pm0.003$ & 95 & 14 & 23 & B & Y & $\delta$ Sct (L,D)\\
272 &      &      & 9777807 & 19:38:54.91 & +46:33:47.7 & 12.728           & 0.345           &                 & 47 & 99 & 60 &   & N & $\gamma$ Dor (D) \\
142 & 4772 &      & 9655407 & 19:37:22.74 & +46:22:53.9 & $12.742\pm0.007$ & $0.366\pm0.013$ & $0.458\pm0.003$ & 96 & 99 & 85 &   & Y & $\delta$ Sct (D)\\
101 & 3464 &      & 9776474 & 19:37:00.19 & +46:31:14.2 & $12.765\pm0.019$ & $0.319\pm0.017$ & $0.428\pm0.004$ & 96 & 99 & 37 &   & Y & $\delta$ Sct (U)\\
152 &      &      & 9532644 & 19:37:26.47 & +46:10:07.4 & 12.786           & 0.350           &                 & 96 & 99 & 47 &   & N & hybrid (U) \\
    &  646 & 1050 & 9836020 & 19:35:43.13 & +46:40:02.7 & $12.799\pm0.004$ & $0.346\pm0.005$ &                 &    & 94 & 91 & Y & N & $\delta$ Sct (U)\\
256 & 8386 &      & 9533489 & 19:38:41.70 & +46:07:21.6 & $12.854\pm0.001$ & $0.346\pm0.002$ &                 & 97 & 97 & 46 & B & N & hybrid (U); 13\\
162 & 5171 &      & 9776816 & 19:37:29.69 & +46:30:39.6 & $12.904\pm0.008$ & $0.333\pm0.015$ & $0.450\pm0.015$ & 96 & 10 & 34 & B & Y & $\gamma$ Dor (U,D)\\
    & 4962 &      & 9897089 & 19:37:25.98 & +46:43:54.2 & $12.945\pm0.008$ & $0.319\pm0.016$ & $0.411\pm0.001$ &    & 98 & 63 & B & N & $\delta$ Sct (D)\\
225 & 7201 &      & 9716947 & 19:38:11.95 & +46:28:02.5 & $12.947\pm0.005$ & $0.383\pm0.008$ &                 & 92 & 99 & 83 & Y & Y & hybrid (U)\\
 27 &  772 &      & 9836073 & 19:35:48.42 & +46:37:29.9 & $12.994\pm0.002$ & $0.341\pm0.002$ &                 & 34 & 99 & 74 & & N & $\delta$ Sct?, misc (D)\\
 94 & 3259 &      & 9716107 & 19:36:56.76 & +46:26:58.0 & $13.073\pm0.012$ & $0.398\pm0.072$ & $0.454\pm0.004$ & 95 & 96 & 79 & Y & Y & $\gamma$ Dor (U)\\
    &      &      & 9410527 & 19:37:30.25 & +45:56:52.3 & 13.110           & 0.397           &                 &    & 99 & 15 & Y & N & $\gamma$ Dor (D)\\
 91 & 3208 &      & 9594100 & 19:36:55.99 & +46:15:18.5 & $13.118\pm0.077$ & $0.431\pm0.003$ &                 & 93 & 99 & 58 & Y & Y & $\gamma$ Dor (U)\\
171 & 5442 &      & 9716523 & 19:37:34.64 & +46:26:01.9 & $13.124\pm0.021$ & $0.346\pm0.023$ & $0.447\pm0.008$ & 96 & 84 &  0 &   & Y & $\gamma$ Dor (D)\\
102 & 3514 &      & 9655151 & 19:37:01.18 & +46:22:59.4 & $13.290\pm0.005$ & $0.404\pm0.016$ & $0.504\pm0.004$ & 97 & 95 & 65 & B & Y & $\gamma$ Dor (U)\\
137 & 4724 &      & 9655399 & 19:37:22.00 & +46:20:50.5 & $13.297\pm0.002$ & $0.381\pm0.001$ & $0.495\pm0.004$ & 97 & 87 & 21 & B & Y & $\gamma$ Dor\\
 51 & 1841 &      & 9654789 & 19:36:28.80 & +46:18:39.2 & $13.333\pm0.007$ & $0.392\pm0.023$ & $0.513\pm0.006$ & 94 & 98 & 61 & B & Y & $\gamma$ Dor (U,D)\\%
 79 &      &      & 9594007 & 19:36:48.86 & +46:14:29.4 & 13.386           & 0.371           &                 & 96 & 91 &  0 &   & Y & $\gamma$ Dor; 14 \\
131 & 4505 &      & 9716350 & 19:37:17.98 & +46:27:45.2 & $13.387\pm0.009$ & $0.392\pm0.013$ & $0.514\pm0.020$ & 95 & 94 &  0 & Y & Y & $\gamma$ Dor; 15\\
209 & 6897 &      & 9594915 & 19:38:02.88 & +46:17:22.5 & $13.639\pm0.010$ & $0.387\pm0.009$ & $0.522\pm0.007$ &  0 & 91 & 24 & & Y & $\delta$ Sct (vC,L); 16
\enddata
\tablenotetext{a}{S71: \citet{sanders6811}; J13: \citet{janes}; UCAC4: \citet{ucac4}, \citet{dias14}; K13: \citet{khar13}.}
\tablenotetext{b}{Full {\it Tycho} ID is TYC 3556-XXXXX-1.}
\tablenotetext{c}{Variability identifications from B: \citet{balona}; D:
  \citet{deboss}; T: \citet{tka}; U: \citet{uytter}; vC: \citet{vanc}, L: \citet{luo}. Square brackets indicate earlier identifications we believe are incorrect. 1) binary? 2) SB1? On blue hook? 3) [misc. (D)]. 4) HIP96532. 5) [$\beta$ Cep (D)]; long P binary. 6) [misc./rot. (D)]. 7) $\gamma$ Dor (T), SPB (D). 8) [$\beta$ Cep (D)]. 9) SB1. 10) Redward of MS. 11) Suspected binary? 12) EB. 13) EB ($P=197.146$ d). 14) [SPB (D)]. 15) [rotation/activity (U)]. 16) faint; nonmember?}
\tablenotetext{d}{Misidentified as S170 in SIMBAD.}
\tablenotetext{e}{Also identified as proper motion member in \citet{baum}.}
\label{pulsate}
\end{deluxetable*}
\end{turnpage}

\begin{turnpage}
\begin{deluxetable*}{rrrcccccccccccl}
\tablewidth{0pc}
\tabletypesize{\scriptsize}
\tablecaption{Miscellaneous Bright Stars in the NGC 6811 Field}
\tablehead{\multicolumn{4}{c}{Identifications\tablenotemark{a}} & \multicolumn{2}{c}{Position}
  & \multicolumn{3}{c}{Photometry} & \multicolumn{5}{c}{Membership\tablenotemark{a}} & 
  \colhead{Notes\tablenotemark{c}} \\
\colhead{S71} & \colhead{J13} & \colhead{Tycho\tablenotemark{b}} & \colhead{KIC} & \colhead{$\alpha$ (2000)} & \colhead{$\delta$ (2000)} & \colhead{$V$} & \colhead{$B-V$} & \colhead{$V-I$} & \colhead{S71} & \colhead{UCAC4} & \colhead{K13} & \colhead{RV} & \colhead{Pos} & }
\startdata
247 & 7930 & 3228 & 9777532 & 19:38:31.06 & +46:31:34.1 & $10.894\pm0.010$ & $0.271\pm0.007$ & $0.352\pm0.002$ & 97 & 99 & 50 & Y & N & 1\\
172 & 5459 & 1874 & 9655543 & 19:37:34.98 & +46:20:53.8 & $11.375\pm0.015$ & $0.395\pm0.008$ & $0.514\pm0.002$ & 97 & 99 & 67\tablenotemark{d} & B & Y & 2\\
110 & 3699 & 1396 & 9655187 & 19:37:04.21 & +46:18:07.8 & $11.499\pm0.009$ & $0.260\pm0.009$ & $0.294\pm0.003$ & 97 & 99 & 97 & B & Y & EB; 3\\
161 & 5148 & 1414 & 9716465 & 19:37:29.26 & +46:29:21.6 & $11.536\pm0.008$ & $0.333\pm0.008$ & $0.462\pm0.002$ & 64 & 99 & 40 & B & Y & 4\\
188 & 5968 &  882 & 9716628 & 19:37:43.64 & +46:29:50.5 & $11.681\pm0.003$ & $0.242\pm0.004$ & $0.336\pm0.001$ & 97 & 99 & 30 & B & Y & 4, 5\\
 77 & 2772 & 1100 & 9655005 & 19:36:48.16 & +46:19:38.0 & $11.883\pm0.008$ & $0.283\pm0.011$ & $0.316\pm0.003$ & 97 & 99 & 85 & B & Y & 4, 5, 6\\
159 & 5112 &      & 9716456 & 19:37:28.53 & +46:24:18.3 & $11.970\pm0.012$ & $0.245\pm0.015$ & $0.316\pm0.003$ & 97 & 99 & 89 & B & Y & EB, 7 \\
134 & 4602 & 1352 & 9655382 & 19:37:19.77 & +46:20:55.3 & $11.992\pm0.008$ & $0.257\pm0.009$ & $0.307\pm0.002$ & 88 & 99 & 30 & Y & Y & 4, 5 \\
189 & 6025 & 2272 & 9716637 & 19:37:44.78 & +46:25:01.0 & $12.111\pm0.006$ & $0.241\pm0.008$ & $0.299\pm0.003$ & 97 & 99 & 99 & B & Y & 8\\
103 & 3534 &      & 9655155 & 19:37:01.59 & +46:20:25.7 & $12.158\pm0.011$ & $0.275\pm0.004$ & $0.347\pm0.003$ & 90 & 99 & 37 &   & Y & 5\\
 46 & 1457 & 2536 & 9531823 & 19:36:18.60 & +46:09:43.5 & $12.266\pm0.005$ & $0.157\pm0.007$ &                 & 66 & 99 & 16 & B & N & 9, 10\\
118 & 4151 &      & 9532445 & 19:37:12.12 & +46:10:13.0 & $12.519\pm0.005$ & $0.309\pm0.006$ &                 & 96 & 99 & 65 & Y & N & 5\\
100 & 3453 &      & 9716154 & 19:37:00.03 & +46:25:21.1 & $12.600\pm0.022$ & $0.605\pm0.009$ & $0.691\pm0.004$ & 57 & 99 & 58 & B & Y & 9, 11\\
123 & 4284 &      & 9655312 & 19:37:14.26 & +46:18:57.2 & $12.741\pm0.011$ & $0.328\pm0.011$ & $0.386\pm0.003$ & 97 & 99 & 24 & Y & Y & 4, 5\\
158 & 5074 &      & 9655471 & 19:37:27.91 & +46:22:26.6 & $13.600\pm0.003$ & $0.505\pm0.001$ & $0.636\pm0.004$ & 96 & 95 & 46 & Y & Y & 9\\
 & 8746 & & 9656397 & 19:38:50.23 & +46:23:05.0 & $13.953\pm0.007$ & $0.575\pm0.018$ & $0.654\pm0.007$ & & 98 &    & Y & N & EB, 12\\
 & 3418 & & 9655129 & 19:36:59.55 & +46:22:26.4 & $14.054\pm0.032$ & $0.810\pm0.012$ & $0.923\pm0.004$ & & 72 &  1 & & Y & EB, 13 \\
 & 6630 & & 9837544 & 19:37:56.73 & +46:40:34.1 & $15.463\pm0.003$ & $0.988\pm0.003$ & $1.098\pm0.006$ & & 94 & 77 & & N & EB, 14
\enddata
\tablenotetext{a}{S71: \citet{sanders6811}; J13: \citet{janes}; UCAC4: \citet{ucac4}, \citet{dias14}; K13: \citet{khar13}.}
\tablenotetext{b}{Full {\it Tycho} ID is TYC 3556-XXXXX-1.}
\tablenotetext{c}{Variability identifications from B: \citet{balona}; D:
  \citet{deboss}; U: \citet{uytter}; vC: \citet{vanc}. 1) rotation/activity (U). 2) HIP96538. 3) $P = 4.4177$ d. 4) misc. (D). 5) rot. (B). 6) 1.5d variations. 7) $P=1.81329$ d. 8) no obvious pulsation. 9) act. (D). 10) odd light curve. 11) nonmember or triple? 12) $P = 204.471$ d. 13) $P = 2.74405$ d; nonmember? 14) $P = 71.6619$ d.}
\tablenotetext{d}{Also identified as proper motion member in \citet{baum}.}
\label{misc}
\end{deluxetable*}
\end{turnpage}
\end{document}